\documentclass[journal,10pt]{IEEEtran}

\usepackage{mathrsfs}
\usepackage{cite}
\usepackage{epsfig}
\usepackage{epsf}
\usepackage{graphics}
\usepackage{subfig}
\usepackage{algorithm}
\usepackage{algorithmicx}
\usepackage{mathtools}
\usepackage{algpseudocode}
\usepackage{graphicx}

\usepackage{color}
\usepackage{url}
\usepackage{hyperref}
\usepackage{flushend}
\usepackage{epstopdf}
\usepackage{cite}
\usepackage{graphicx}
\usepackage{textcomp}
\usepackage{bm}
\usepackage{xcolor}
\def\BibTeX{{\rm B\kern-.05em{\sc i\kern-.025em b}\kern-.08em
    T\kern-.1667em\lower.7ex\hbox{E}\kern-.125emX}}
\usepackage{amsmath,graphicx,amssymb,amsfonts}
\usepackage{algorithm}
\usepackage{algpseudocode}
\def\BibTeX{{\rm B\kern-.05em{\sc i\kern-.025em b}\kern-.08em
    T\kern-.1667em\lower.7ex\hbox{E}\kern-.125emX}}
\definecolor{brown}{rgb}{0.59, 0.29, 0.0}
\begin{document}
\title{Joint Transmission Scheme and Coded Content Placement in Cluster-centric UAV-aided Cellular~Networks}

\author{Zohreh HajiAkhondi-Meybodi,~\IEEEmembership{Student Member,~IEEE}, Arash Mohammadi,~\IEEEmembership{Senior Member,~IEEE}, \\ Jamshid Abouei,~\IEEEmembership{Senior Member,~IEEE}, Ming Hou,~\IEEEmembership{Senior Member,~IEEE}, and~Konstantinos~N.~Plataniotis,~\IEEEmembership{Fellow,~IEEE}% <-this % stops a space
\thanks{Z. HajiAkhondi-Meybodi is with Electrical and Computer Engineering (ECE), Concordia University, Montreal, Canada. (E-mail: z\_hajiak@encs.concordia.ca). A. Mohammadi (corresponding author) is with Concordia Institute of Information Systems Engineering (CIISE), Concordia University, Montreal, Canada. (P: +1 (514) 848-2424 ext. 2712 F: +1 (514) 848-3171, E-mail: arash.mohammadi@concordia.ca). J. Abouei was with the Department of Electrical and Computer Engineering, University of Toronto, Toronto, Canada. He is now with the Department of Electrical Engineering, Yazd University, Yazd 89195-741, Iran (E-mail: abouei@yazd.ac.ir). M. Hou is with Defence Research and Development Canada (DRDC), Ottawa, Toronto, ON, M2K 3C9, Canada. (E-mail: ming.hou@drdc-rddc.gc.ca). K.~N.~Plataniotis is with Electrical and Computer Engineering (ECE), University of Toronto, Toronto, Canada. (E-mail: kostas@ece.utoronto.ca).}

\thanks{This Project was partially supported by Department of National Defence's Innovation for Defence Excellence \& Security (IDEaS)
program, Canada.}}% <-this % stops a space

\markboth{}%Internet of Things Journal,~Vol.~XX, No.~X, XX~2021}%
{Hajiakhondi \MakeLowercase{\textit{et al.}}:DQLEL: Deep Q-Learning for Energy-Optimized LoS/NLoS UWB Node Selection}
\maketitle
\begin{abstract}
Recently, as a consequence of the COVID-19 pandemic, dependence on telecommunication for remote learning/working and telemedicine has significantly increased. In this context, preserving high Quality of Service (QoS) and maintaining low latency communication are of paramount importance. In cellular networks, incorporation of Unmanned Aerial Vehicles (UAVs) can result in enhanced connectivity for outdoor users due to the high probability of establishing Line of Sight (LoS) links. The UAV's limited battery life and its signal attenuation in indoor areas, however, make it inefficient to manage users' requests in indoor environments. %Despite all the researches on the UAV-aided cellular networks, there is no efficient infrastructure that can be beneficial for indoor users, as well. 
Referred to as the Cluster-centric and Coded UAV-aided Femtocaching (CCUF) framework, the network's coverage in both indoor and outdoor environments increases by considering a two-phase clustering framework for FAPs' formation and UAVs' deployment. Our first objective is to increase the content diversity. In this context, we propose a coded content placement in a cluster-centric cellular network, which is integrated with the Coordinated Multi-Point (CoMP) approach to mitigate the inter-cell interference in edge areas. Then, we compute, experimentally, the number of coded contents to be stored in each caching node to increase the cache-hit-ratio, Signal-to-Interference-plus-Noise Ratio (SINR), and cache diversity and decrease the users' access delay and cache redundancy for different content popularity profiles. Capitalizing on clustering, our second objective is to assign the best caching node to indoor/outdoor users for managing their requests. In this regard, we define the movement speed of ground users as the decision metric of the transmission scheme for serving outdoor users' requests to avoid frequent handovers between FAPs and increase the battery life of UAVs. Simulation results illustrate that the proposed CCUF implementation increases the cache-hit-ratio, SINR, and cache diversity and decrease the users' access delay, cache redundancy and UAVs’ energy consumption.
\end{abstract}


\IEEEpeerreviewmaketitle
%OOOOOOOOOOOOOOOOOOOOOOOOOOOOOOOOOOOOOOOOOOOOOOOOOOOOOOOOO
\section{Introduction} \label{Introduction}
%OOOOOOOOOOOOOOOOOOOOOOOOOOOOOOOOOOOOOOOOOOOOOOOOOOOOOOOOO
\vspace{-0.05in}
\IEEEPARstart{A}{s} a consequence of the COVID-19 pandemic, dependence on telemedicine and remote learning/working has significantly increased due to the exponential rise in the demand for in-home care, remote working, schooling, and remote reporting~\cite{Chamola2020}. Caching has emerged as a promising solution to maintain low latency communication and mitigate the network's traffic over the backhaul. This, in turn, improves the Quality of Service (QoS) by storing the most popular multimedia content close to the end-users~\cite{Hajiakhondi2019, Hajiakhondi2020}. Due to the limited storage of caching nodes, however, it is critical to increase their content diversity, therefore, storing more contents in caching nodes. Recently, Unmanned Aerial Vehicle (UAV)-based caching has gained significant attention from both industry and academia~\cite{Li2019,Jiang2019, Cheng2019, Zhao2018, Sharma2019, Wang2019, Liu2019,Samir2019, Li2020_2,Chen2019, Yang2019,Ji2019, Zhang2018,Suman2018,Hu2018_2}, due to its high-quality Line of Sight (LoS) links. Although the enhanced connectivity that comes by using UAVs will improve the QoS in outdoor areas, indoor penetration loss and deep shadow fading caused by building walls significantly attenuate the UAV's signals in indoor environments degrading the network's overall QoS~\cite{Lee2019}.  Capitalizing on the above discussion, the paper focuses on the issues of  content diversity and transmission scheme of indoor/outdoor users. In this context and in line with advancements of 5G networks, the paper targets coupling UAVs as aerial caching nodes with Femto Access Points (FAPs)~\cite{Avanzato2020}. 
In what follows, to understand the state-of-the-art in this area and seek potential solutions, we first review the relevant literature.

\vspace{.025in}
\noindent
\textbf{Related Work:} The main objective of UAV-aided cellular networks is to bring multimedia data closer to ground users and simultaneously improve users' QoS and the network's Quality of Experience (QoE). If  the  requested  content  can be found  in  the  storage of one of the available caching nodes, this  request would be served directly and cache-hit occurs; otherwise, it is  known as  a cache-miss. Due to the large size of multimedia contents, however, it is not feasible to store all contents in the storage of caching nodes. To increase content diversity, coded caching strategies~\cite{Recayte2018,Ko2019} 
have received remarkable attention lately. In coded caching strategies, only specific segments of the most popular multimedia contents are stored in the caching nodes. Early works on coded femtocaching such as Reference~\cite{Shanmugam2013}, however, considered \textit{homogeneous} networks, where the same segments of the most popular multimedia contents are stored in different caching nodes. To boost the content diversity, 
the main focus of recent researchers has been shifted to the cluster-centric networks~\cite{Chen2017_2} as a \textit{heterogeneous} infrastructure, where distinct segments are stored in neighboring caching nodes.

Cluster-centric cellular networks provide several benefits, such as increased content/cache diversity, which in turn leads to an increase in the number of requests managed by the  caching nodes. However, this comes with the cost of experiencing inter-cell interference, especially for cell-edge users. To mitigate the inter-cell interference and improve the throughput of the cell-edge users, content caching in Coordinated Multi-Point (CoMP)-integrated networks~\cite{Liu2014,Yu2019,Chen2017_2,Lin2020,Li2017,Lin2019} has been studied in recent years. For instance, Chen \textit{et al.}~\cite{Chen2017_2} developed two transmission schemes, namely Joint Transmission (JT) and Parallel Transmission (PT), which are selected based on the popularity of the requested content. Alternatively, Lin \textit{et al.}~\cite{Lin2020} proposed a cluster-centric cellular network applying the CoMP technique based on the users' link quality, where cell-core and cell-edge users are served through Single Transmission (ST) and JT, respectively. \textit{Despite all the researches on the cluster-centric cellular networks, there is no framework to determine how different segments can be cached to increase the data availability in a UAV-aided cluster-centric cellular network to increase content diversity}. The paper addresses this~gap.

In addition to content diversity, one of the main challenges in UAV-aided cellular networks is to optimally assign caching nodes (UAVs or FAPs) to ground users to efficiently serve their requests~\cite{Athukoralage2016, Zhu2020_2}. There are several QoS and QoE metrics that can be considered as the decision criteria for Access Point (AP) selection, including users' latency; traffic load and energy consumption of APs; users' link quality; handover rate, and; Signal-to-Interference-plus-Noise Ratio (SINR). For instance, Athukoralage \textit{et al.}~\cite{Athukoralage2016} considered an AP selection framework, where ground users are supported by UAV or WiFi APs. In this work, the users' link quality is utilized to balance the load between UAVs and WiFi APs. Zhu \textit{et al.}~\cite{Zhu2020_2} proposed a game theory-based AP selection scheme, where the probability of packet collision is used to select the optimal AP among all possible UAVs and Base Stations (BSs). In our previous work~\cite{Hajiakhondi2021}, we introduced the Convolutional Neural Network (CNN) and Q-Network-based Connection Scheduling (CQN-CS) framework with the application to the UAV-aided cellular networks. In that work, without considering the content diversity, ground users were autonomously trained to determine the optimal caching node, i.e., UAVs or FAPs. Although we minimized the users' access delay by maintaining a trade-off between the energy consumption of UAVs and the occurrence of handovers between FAPs, there are still key challenges ahead. Our previous work is just efficient for serving ground users in outdoor environments as the UAV's signal attenuation in indoor environments is not factored in. 
The wide transmission range of UAVs and the high probability of establishing LoS links  provide several advantages, including the ability to manage the majority of ground users' requests, which leads to improved coverage in outdoor environments. Due to the limited battery life of UAVs, however, requests that are handled by UAVs should be controlled. Another challenge is the handover phenomenon, which can be frequently triggered by FAPs if the ground user moves rapidly and leaves the current position. \textit{To date, limited research has been performed on UAV-aided cellular networks to provide high QoS for ground users in both indoor and outdoor environments.} The paper also addresses this gap.

\vspace{.025in}
\noindent
\textbf{Contribution:} In this paper, we consider an integrated UAV-aided and cluster-centric cellular network to serve ground users positioned in both indoor and outdoor environments. Our first objective is to increase the content diversity that can be accessed via caching nodes. The second goal is to introduce different transmission schemes for indoor/outdoor users to improve the achievable QoS in terms of the users' access delay and decrease the energy consumption of UAVs. Referred to as the Cluster-centric and Coded UAV-aided Femtocaching (CCUF) framework, the network's coverage in both indoor and outdoor environments increases by considering a two-phase clustering approach for FAPs' formation and UAVs' deployment. In summary, the paper makes the following key contributions:
\begin{itemize}
\item Due to the UAV's signal attenuation in indoor environments, we consider two different indoor and outdoor caching service scenarios for the proposed CCUF framework. More precisely, the indoor area is covered by FAPs, equipped with extra storage and supported by CoMP technology. The outdoor area, however, is supported by coupled UAVs and FAPs depending on the movement speed of ground users.
\item To access a large number of content during the movement of ground users, a two-phase clustering  approach is considered: $(i)$ The whole network (both indoor and outdoor areas) is partitioned into sub-networks called inter-clusters, which is defined for content placement in FAPs. We show that based on this strategy, the ground users can acquire more segments during their movements, and; $(ii)$ For UAVs formation, the outdoor environment is partitioned into intra-clusters via a  $K$-means clustering  algorithm, each covered by a UAV.
\item To  the  best  of  our knowledge, despite all the research conducted in this field, there is no placement strategy to determine how distinct segments of popular content should be distributed in different caching nodes. Towards this goal, we consider a cluster-centric cellular network, where multimedia contents are classified into three categories, including popular, mediocre, and non-popular contents. While the popular contents are stored completely, distinct segments of mediocre ones are determined according to the proposed framework to be stored in the storage of neighboring FAPs.  We also determine the best number of coded/uncoded contents in each caching node to increase the cache-hit-ratio, SINR, and cache diversity while decreasing users' access delay and cache redundancy for different content popularity profiles.
\end{itemize}
The effectiveness of the proposed CCUF framework is evaluated through simulation studies in both indoor and outdoor environments in terms of cache-hit-ratio, users' access delay, SINR, cache diversity, cache redundancy, and energy consumption of UAVs. According to the simulation results, to improve the aforementioned metrics for different content popularity profiles, we investigate the best number of coded contents in each caching node.

The remainder of the paper is organized as follows: In Section~\ref{sec:sys}, the network's model is described and the main assumptions required for the implementation of the proposed framework are introduced. Section~\ref{sec:Prop} presents the proposed CCUF scheme. Simulation results are presented in Section~\ref{sec:sim}. Finally, Section~\ref{sec:con} concludes the paper.

%%%%%%%%%%%%%%%%%%%%%%%%%%%%%%%%%%%%%%%%%
\setlength{\textfloatsep}{0pt}
\begin{figure*} [t]
\centering \includegraphics [scale = 0.24] {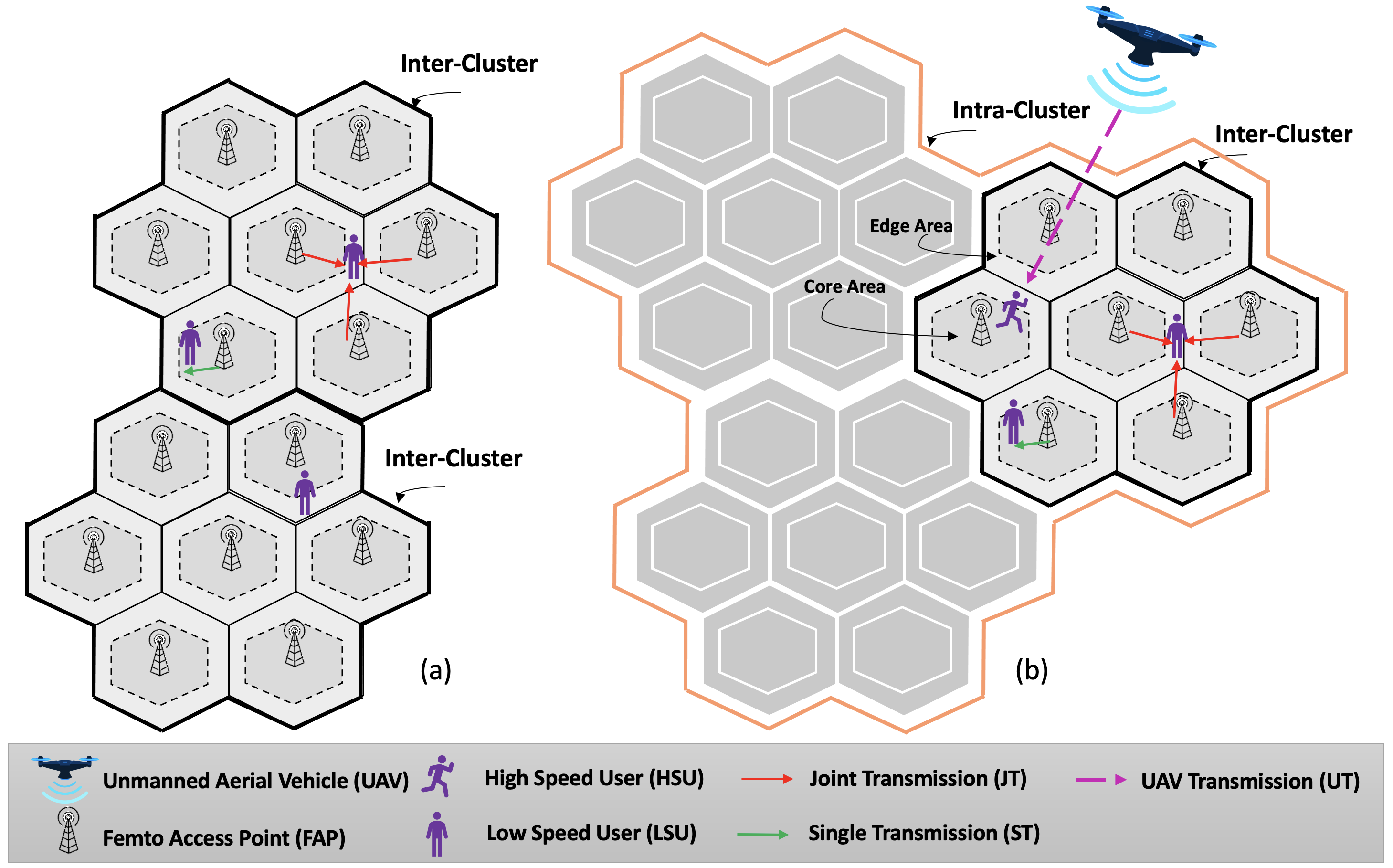}
\caption{\footnotesize A typical structure of the proposed UAV-aided cellular network in (a) the indoor, and (b) the outdoor environments.} \label{sys}
\end{figure*}
%%%%%%%%%%%%%%%%%%%%%%%%%%%%%%%%%%%%%%%%%

%OOOOOOOOOOOOOOOOOOOOOOOOOOOOOOOOOOOOOOOOOOOOOOOOOOOOOOOOO
\section{System Model and Problem Description} \label{sec:sys}
%OOOOOOOOOOOOOOOOOOOOOOOOOOOOOOOOOOOOOOOOOOOOOOOOOOOOOOOOO
We consider a UAV-aided cellular network in a residential area that supports both indoor and outdoor environments (see Fig.~\ref{sys}). There exist $N_f$ number of FAPs, denoted by $f_i$, for ($1 \leq i \leq N_f$), each with the cache size of $C_f$ and transmission range of $R_f$. All FAPs are independently and randomly distributed in the environment following Poisson Point Processes (PPPs). There are also $N_u$ number of UAVs, denoted by $u_{k}$, for ($1 \leq k \leq N_u$), with equal transmission range of $R_u$, and a main server that has access to the whole content and can manage all caching nodes. There are $N_g$ number of ground users, denoted by $GU_j$, for ($1 \leq j \leq N_g$), that move through the network with different velocities. Term  $\upsilon_{j}(t)$ denotes the speed of the ground user $GU_j$ at time slot $t$. When $GU_j$ requests content $c_l$ from a library of $\mathcal{C} = \lbrace c_1, \ldots, c_{N_c} \rbrace$, in which $N_c=|\mathcal{C}|$ is the cardinality of multimedia data in the network, this request should be handled by one of the nearest FAPs or UAVs having some segments of $c_l$. In this work, FAP $f_i$, for ($ 1\leq i \leq N_f$), and UAV $u_{k}$, for ($1 \leq k \leq N_u$), operate in an open access mode, i.e., they can serve any ground user $GU_j$, for ($1 \leq j \leq N_g$), located in their transmission range. To completely download a requested content, a finite time $T$ is required. In the proposed CCUF framework, each content $c_l$ is fragmented into $N_s$ encoded segments, denoted by $c_{ls}$, for ($ 1\leq s \leq N_s$). Without loss of generality, it is assumed that the time $T$ is discretized into $N_s$ time slots with time interval $\delta_t$, i.e., $T = N_s\delta_t$, where $\delta_t$ is large enough for downloading one segment $c_{ls}$.

As it can be seen from Fig.~\ref{sys}, we consider two different indoor and outdoor caching service scenarios for the proposed CCUF framework to improve the network's coverage. As will be described in Subsection~\ref{subsec:TS}, the requests of indoor users are handled through FAPs, while outdoor users are supported by coupled UAVs and FAPs depending on their movement speed. In this regard, we define two clustering approaches, called inter-clusters and intra-clusters, which are used for content placement in FAPs and UAVs' deployment, as discussed below:

\vspace{.025in}
\noindent
\textbf{Inter-Clusters:} As shown in Fig.~\ref{sys}, $N_b < N_f$ number of neighboring FAPs in both indoor and outdoor environments, form a cluster, referred to as the inter-cluster. As it is stated in~\cite{Chen2017_2, Afshang2016}, the main focus of cluster-centric content placement is to place contents in the storage of FAPs, where all FAPs belonging to an inter-cluster are used as an entity (despite conventional femtocaching schemes where each FAP acts as single storage). Therefore, our goal is to determine how different segments of popular files should be distributed in the cache of FAPs belonging to an inter-cluster to increase the content diversity. We construct the inter-clusters based on the following two rules: (i) As will be described shortly in Subsection~\ref{sec:CPP}, all FAPs in the same inter-cluster save \textit{different} segments of mediocre contents, and; (ii) The cached contents of different inter-clusters are the \textit{same}. In addition, all FAPs use the CoMP transmission approach (supporting ST and JT schemes) to mitigate the inter-cell interference in edge areas and manage ground users' requests.

\vspace{.025in}
\noindent
\textbf{Intra-Clusters:} Since the transmission range of FAPs is significantly less than that of a UAV (see Fig.~\ref{sys}(b)), the outdoor area is also divided into several intra-clusters (each intra-cluster is covered by a UAV) based on an unsupervised learning algorithm.
In what follows, we present the content popularity profile and transmission schemes utilized to develop the proposed CCUF framework.

%=====================================================
\subsection{Content Popularity Profile} \label{sec:CPP}
%=====================================================
To account for user's behavior pattern in multimedia services, the popularity of video contents is determined based on the Zipf distribution, where the probability of requesting the $l^{\text{th}}$ file, denoted by $p_l$, is calculated as
\begin{equation}\label{nEq:5}
p_l=\dfrac{l^{-\gamma}}{\sum \limits_{r=1}^{N_c}{r^{-\gamma}}},
\end{equation}
where $\gamma$ and $r$ represent the skewness of the file popularity, and the rank of the file $c_l$, respectively. For notational convenience, we assume that $P[n] \equiv P(n \delta_t)$ denotes the probability of accessing a new segment in time slot $n$, with $n \in \{1, \ldots, N_s\}$. Without loss of generality and to be practical, we investigate the probability distribution of a real multimedia data set, i.e., the YouTube videos trending statistics, following Zipf distribution, i.e., a small part of the contents are requested with a high probability. The majority of contents are not popular, and some contents, are requested moderately. Consequently, in the proposed CCUF framework, we classify multimedia contents into three categories, i.e., popular, mediocre, and non-popular~\cite{Chen2017_2}. To improve content diversity, the storage capacity of FAPs, denoted by $C_f$, is divided into two spaces, where $\alpha$ portion of the storage is allocated to store complete popular contents, i.e., $1 \leq l \leq \lfloor \alpha C_f \rfloor$, where $l=1$ indicates the most popular content. Additionally, $(1-\alpha)$ portion of the cache is assigned to store different parts of the mediocre contents, where $\lfloor \alpha C_f \rfloor +1 \leq l \leq  N_s ( C_f - \lfloor \alpha C_f \rfloor)$. The best value of $\alpha$ is obtained experimentally. The proposed model for identifying different segments to be cached in neighboring FAPs will be discussed later on in Section~\ref{sec:Prop}.

%=====================================================
\subsection{Transmission Scheme}\label{subsec:TS}
%=====================================================
In this subsection, we describe both the connection scheduling (serving by FAPs or UAVs) and the transmission scheme depending on the presence of the ground user in indoor or outdoor environments.

%----------------------------------------------------------------------------------------------------------
\subsubsection{Indoor Environment}
%----------------------------------------------------------------------------------------------------------
The transmitted signal by UAVs, propagating in residential areas, becomes weaker due to the penetration loss and shadow fading effects. It is, therefore, assumed that ground users positioned in indoor areas are only supported by FAPs. In the CoMP-integrated and cluster-centric cellular network and as it can be seen from Fig.~\ref{sys}, there are two regions in each inter-cluster, named cell-edge and cell-core, which are determined based on the long-term averaged SINR values~\cite{Lin2020} to illustrate the quality of a wireless link. In such a case that the ground user $GU_j$ is positioned in the vicinity of the FAP $f_i$, the SINR from $f_i$ to $GU_j$, denoted by $\mathcal{S}_{i,j}$, is obtained as follows
\begin{equation}\label{nEq:6}
\mathcal{S}_{i,j}(t)= \dfrac{P_i |\mathcal{\tilde{H}}_{i,j}(t)|^2 }{I_{\bm{f}_{-i}}(t)+N_0},
\end{equation}
where $P_i$ denotes the transmitted signal power of FAP $f_i$, and $I_{\bm{f}_{-i}}(t)$  represents the interference power from other FAP-ground users, except for the corresponding $f_i$ link. Term $N_0$ represents the noise power related to the additive white Gaussian random variable. Moreover, the path loss and fading channel effects between FAP $f_i$ and ground user $GU_j$ at time slot $t$ is denoted by $\mathcal{\tilde{H}}_{i,j}(t)= \dfrac{h_{i,j}(t)}{\sqrt{\mathcal{L}_{i,j}(t)}}$. In this case, $h_{i,j}(t)$ denotes a complex zero-mean Gaussian random variable with unit standard deviation and $\mathcal{L}_{i,j}(t)$ represents the path loss between FAP $f_i$ and ground user $GU_j$ at time slot $t$, obtained as follows
\begin{equation}\label{nEq:7}
\mathcal{L}_{i,j}(t)=\mathcal{L}_0+10\eta\log \big( d_{i,j}(t) \big)+\chi_{\sigma},
\end{equation}
where $\eta$ is the path loss exponent. Term $\chi_{\sigma}$ indicates the shadowing effect,  which is a zero-mean Gaussian-distributed random variable with standard deviation $\sigma$.  Additionally, $d_{k,j}(t)$ represents the Euclidean distance between FAP $f_i$ and ground user $GU_j$ at time slot $t$. Furthermore, $\mathcal{L}_0=20 \log \left(\dfrac{4\pi f_c d_0}{c} \right) $ is the path loss related to the reference distance $d_0$ where $f_c$ and $c= 3 \times 10^8$ denote the carrier frequency and the light speed, respectively.  Accordingly, the ground user $GU_j$ is marked as the cell-core user connected to FAP $f_i$, if $\overline {\mathcal{S}}_{i,j}(t) > \mathcal{S}_{th}$; otherwise, $GU_j$ is marked as the cell-edge user, where $\mathcal{S}_{th}$ is the SINR threshold.

The transmission scheme in the proposed CoMP-integrated and cluster-centric cellular network is determined based on two metrics; (i) The popularity of the requested content, described in Subsection~\ref{sec:CPP}, and; (ii) The link quality of the ground user in the cell, i.e., cell-core or cell-edge. The following two different transmission schemes are utilized for the development of the proposed CCUF framework:
\begin{itemize}
\item \textbf{Single Transmission (ST):} In this case, the requested file $c_l$, for $(1 \leq l \leq \lfloor \alpha C_f \rfloor)$, is a popular content, and the ground user $GU_j$ is marked as the cell-core of FAP $f_i$, i.e., $\overline{\mathcal{S}}_{i,j}(t) > \mathcal{S}_{th}$. It means that the content is completely cached into the storage of FAP $f_i$ and the high-quality link can be established between FAP $f_i$ and ground user $GU_j$. Consequently, this request is served only by the corresponding FAP $f_i$. Moreover, if the requested content belongs to the mediocre category, i.e., $\lfloor \alpha C_f \rfloor +1 \leq l \leq  N_s ( C_f - \lfloor \alpha C_f \rfloor)$, this request is served according to the ST scheme regardless of the user's link quality since each FAP has a different segment of the mediocre content.
\item \textbf{Joint Transmission (JT):} In this transmission scheme, the requested file $c_l$ is a popular content, i.e., $1 \leq l \leq \lfloor \alpha C_f \rfloor$. Consequently, all FAPs have the same complete file. The ground user $GU_j$, however, is marked as the cell-edge of FAP $f_i$, i.e., $\overline{\mathcal{S}}_{i,j}(t) \leq \mathcal{S}_{th}$. Therefore, the link quality between FAP $f_i$ and the ground user $GU_j$ is not good enough. In order to improve the reliability of content delivery, the corresponding content will be jointly transmitted by several FAPs in its inter-cluster. As it can be seen from Fig.~\ref{sys}, neighboring FAPs in an inter-cluster collaboratively serve cell-edge ground users based on the JT scheme, which is shown by the red color.
\end{itemize}

%----------------------------------------------------------------------------------------------------------
\subsubsection{Outdoor Environment}
%----------------------------------------------------------------------------------------------------------
As it can be seen from Fig.~\ref{sys}, outdoor areas are covered by both  UAVs and FAPs. Therefore, outdoor users are classified based on their velocity into the following two categories:
\begin{itemize}
\item \textbf{Low Speed Users (LSUs):} If the speed of ground user $GU_j$, denoted by $\upsilon_{j}(t)$, is less than a predefined threshold $\upsilon_{\rm{th}}$, this user is managed by inter-clusters (FAPs). Therefore, the transmission scheme of LSUs is completely the same as the indoor users, described in Subsection~\ref{subsec:TS}-1.
\item \textbf{High Speed Users (HSUs):} In this case, the speed of ground user $GU_j$ is equal or more than $\upsilon_{\rm{th}}$. Therefore, this request should be served by a UAV covering the corresponding intra-cluster.
\end{itemize}
This completes our discussion on the content popularity profile and transmission schemes. Next, we develop the CCUF framework.

%OOOOOOOOOOOOOOOOOOOOOOOOOOOOOOOOOOOOOOOOOOOOOOOOOOOOOOOOO
\section{The CCUF Framework} \label{sec:Prop}
%OOOOOOOOOOOOOOOOOOOOOOOOOOOOOOOOOOOOOOOOOOOOOOOOOOOOOOOOO
In conventional femtocaching schemes, it is a common assumption that all caching nodes store the same most popular contents. This assumption is acceptable in static femtocaching models, in which users are stationary or move with a low velocity. With the focus on a dynamic femtocaching network, in which users can move based on the random walk model, storing distinct content in neighboring FAPs leads to increasing the number of requests served by caching nodes~\cite{SZhang2018}. Despite recent researches on cluster-centric cellular networks, there is no framework to determine how different segments should be stored to increase content diversity. Toward this goal, we propose the CCUF framework, which is an efficient content placement strategy for the network model introduced in Section~\ref{sec:sys}.  The proposed CCUF framework is implemented based on the steps presented in the following subsections.

%=====================================================
\subsection{Content Caching for FAPs and UAVs}
%=====================================================
Identifying the best multimedia content to be stored in the storage of caching nodes leads to a reduction in the users' latency. It is commonly assumed~\cite{Shanmugam2013, Hajiakhondi2019} that the total users' access delay is determined according to the availability of  the required content in the nearby caching nodes. Based on this assumption, in scenarios where the requested content can be served by caching nodes, the cache-hit occurs and the ground user will experience no delay; otherwise, the request is served by the main server resulting in a cache-miss. Inspired from~\cite{Xue2019}, we relax the above assumption and express the actual users' access delay as a function of the content popularity profile and link's quality. In this regard, we propose two optimization models for content placement in both FAPs and UAVs to minimize the users' latency, which are similar in nature. Toward this goal, we first describe the delay that ground users experience when served by FAPs and UAVs.

%----------------------------------------------------------------------------------------------------------
\subsubsection{UAVs' Content Placement}
%----------------------------------------------------------------------------------------------------------
Serving requests via UAVs leads to establishing air-to-ground links from UAVs to ground users. Due to the obstacles in outdoor environments, the transmitted signal from UAVs is attenuated. To be practical, we consider both LoS and Non-LoS (NLoS) path losses from UAV $u_k$ to ground user $GU_j$ at time slot $t$ as follows~\cite{Jiang2019}
\begin{eqnarray}
\!\!\!\!\mathcal{L}_{k,j}^{(LoS)}(t) &\!\!\!\!=\!\!& \mathcal{L}_0+10\eta^{(LoS)}\log(d_{k,j}(t))+\chi_{\sigma}^{(LoS)},\label{nEq:8}\\
\!\!\!\!\mathcal{L}_{k,j}^{(NLoS)}(t) &\!\!\!\!=\!\!& \mathcal{L}_0+10\eta^{(NLoS)}\log(d_{k,j}(t))+\chi_{\sigma}^{(NLoS)},\label{nEq:9}
\end{eqnarray}
where $\mathcal{L}_0=20 \log \left(\dfrac{4\pi f_c d_0}{c} \right) $ denotes the reference path loss in distance $d_0$, and $d_{k,j}(t)$ is the Euclidean distance between UAV $u_k$ and the ground user $GU_j$ at time slot $t$. In addition, $\eta^{(LoS)}$, $\eta^{(NLoS)}$,  $\chi_{\sigma}^{(LoS)}$ and $\chi_{\sigma}^{(NLoS)}$ indicate the LoS and NLoS path loss exponents and the corresponding shadowing effects, respectively. Consequently, the average path loss, denoted by $\mathcal{\overline L}_{k,j}(t)$, is  given  by
\begin{equation}\label{nEq:10}
\mathcal{\overline L}_{k,j}(t)=p^{(LoS)}_{k,j}(t) \mathcal{\overline L}_{k,j}^{(LoS)}(t) + (1-p^{(LoS)}_{k,j}(t))\mathcal{\overline L}_{k,j}^{(NLoS)}(t),
\end{equation}
where $p^{(LoS)}_{k,j}(t)$ is the probability of establishing LoS link between UAV $u_k$ and  ground user $GU_j$ at time slot $t$, obtained as~\cite{Chen2019}
\begin{equation}\label{nEq:11}
p^{(LoS)}_{k,j}(t)=\left(1+\vartheta \exp{\left(-\zeta [\phi_{k,j}(t)-\vartheta]\right)}\right)^{-1},
\end{equation}
where $\vartheta$ and $\zeta$ are constant parameters, depending on the rural and urban areas. Moreover, $\phi_{k,j}(t)= \sin^{-1}{\left(\dfrac{h_k}{d_{k,j}(t)}\right)}$ is the elevation angle between UAV $u_k$ and the ground user $GU_j$, and $h_k$ is the UAV's altitude. Without loss of generality, altitude $h_k$ is assumed to be a fixed value over the  hovering time. If the requested content cannot be found in the storage of UAVs,  additional ground-to-air connection is required to provide UAVs with the requested content through the main server. Similarly, the average path loss of the main server-to-UAV $u_k$ link is calculated as
\begin{equation}\label{nEq:12}
\mathcal{\overline L}_{m,k}(t)=p^{(LoS)}_{m,k}(t) \mathcal{L}_{m,k}^{(LoS)}(t) + (1-p^{(LoS)}_{m,k}(t))\mathcal{L}_{m,k}^{(NLoS)}(t),
\end{equation}
where $\mathcal{L}_{m,k}^{(LoS)}(t)=d_{m,k}^{-\varpi}(t)$ and  $\mathcal{L}_{m,k}^{(NLoS)}(t)=\psi \mathcal{L}_{m,k}^{(LoS)}(t)$, in which $d_{m,k}(t)$ denotes the distance between the main server and UAV $u_k$. Furthermore, $\varpi$ and $\psi$ denote the LoS and NLoS path loss exponents, respectively~\cite{Jiang2019}.

As stated previously, another parameter that has a great impact on the users' access delay is the presence of the requested content in the caching node, depending on the content popularity profile. Therefore, the cache-hit and the cache-miss probability through serving by UAV $u_k$ at time slot $t$, denoted by $p_u^{(h)}(t)$ and $p_u^{(m)}(t)$, respectively, are expressed~as
\begin{eqnarray}
p_u^{(h)}(t) &=& \sum_{l \in C_u} p_l (t)	\leq 1,\label{nEq:13}\\
p_u^{(m)}(t) &=& 1-p_u^{(h)}(t),\label{nEq:14}
\end{eqnarray}
where $C_u$ denotes the cache size of UAV $u_k$, which is assumed to be the same for all UAVs.  Consequently, the users' access delay through UAVs is expressed as
\begin{equation}\label{nEq:15}
\mathcal{D}_u(t)=p_u^{(h)}(t) \mathcal{D}_u^{(h)}(t)+p_u^{(m)}(t) \mathcal{D}_u^{(m)}(t),
\end{equation}
where $\mathcal{D}_u^{(h)}(t)$ and $\mathcal{D}_u^{(m)}(t)$ represent the cache-hit and the cache-miss delays, respectively, calculated as follows~\cite{Chen2019}
\begin{eqnarray}
\!\!\!\!\mathcal{D}_u^{(h)}(t) &\!\!=\!\!& \dfrac{L_c}{R_{k,j}} = L_c \log^{-1}\left (1+\dfrac{P_k 10^{\mathcal{\overline L}_{k,j}(t)/10} }{I_k(t,\bm{u}_{-k})+N_0} \right),\label{nEq:16}\\
\!\!\!\!\mathcal{D}_u^{(m)}(t) &\!\!=\!\!& \underbrace{ L_c \log^{-1} \left (1+\dfrac{P_k 10^{\mathcal{\overline L}_{m,k}(t)/10}}{I_k(t,\bm{u}_{-k})+N_0} \right)}_{\triangleq\mathbf{L_{MU}}} + \nonumber\\
&&\underbrace{ L_c \log^{-1} \left (1+\dfrac{P_k 10^{\mathcal{\overline L}_{k,j}(t)/10}}{I_k(t,\bm{u}_{-k})+N_0} \right)}_{\triangleq\mathbf{L_{UG}}},\label{nEq:17}
\end{eqnarray}
where $L_c$ and $R_{k,j}$ represent the file size of $c_l$ and the transmission data rate from UAV $u_k$ to $GU_j$. Furthermore, $P_k$ and $I_k(t,\bm{u}_{-k})$ denote the transmission power of UAV $u_k$ and the interference power caused by other UAV-user links for the transmission link between $u_k$ and $GU_j$, respectively. Note that when the cache-miss happens, the content should be first provided for the UAV by the main server. Therefore, $L_{MU}$ and $L_{UG}$ in Eq.~\eqref{nEq:17} represent the users' access delay related to the main server-UAV and UAV-ground user links, respectively.

Given users' access delay through UAVs, the goal is to place contents in the storage of UAVs to minimize the users' access delay in Eq.~\eqref{nEq:15}. Due to the large coverage area of UAVs, it is not feasible for ground users to move through areas supported by different UAVs frequently. Therefore, we assume that contents (either popular or mediocre ones) are cached completely in the storage of UAVs. With the aim of minimizing users' access delay,  the cached contents are selected as the solution of the following optimization problem: 
\begin{eqnarray}{}\label{nEq:18}
\min \limits_{\substack{\mathrm{x}_l}}~~ \sum \limits_{l=1}^{N_c} \Big ( \sum \limits_{j=1}^{N_g}
p_l^{(j)}(t)  \mathcal{D}_{u}^{(j)} (t)\Big ) \mathrm{x}_l  \label{nEq:18}\\ \nonumber
\text{s.t.}~~ \textbf{C1.}~~  \mathrm{x}_l \in \{0,1\},
\\ \nonumber
~~~~~~~~~~~~~~~~ \textbf{C2.}~~ \sum \limits_{l=1}^{N_c} (1-\mathrm{x}_l) \leq C_u,
\end{eqnarray}
where $p_l^{(j)}(t)$ denotes the probability of requesting content $c_l$ by the ground user $GU_j$ at time slot $t$, which is obtained according to the request history of ground user $GU_j$~\cite{Hajiakhondi2019}. Furthermore, $\mathcal{D}_{u}^{(j)}(t)$ is the delay that the ground user $GU_j$ may experience, which is calculated based on Eq.~\eqref{nEq:15}. In the constraint \textbf{C1}, $\mathrm{{x}}_{l}$ is an indicator variable, which is equal to $0$ when content $c_{l}$ exists in the cache of UAV $u_k$. Moreover, the constraint~\textbf{C2} represents that the total contents cached in the storage of $u_k$ should not exceed its storage capacity of $u_k$.

%----------------------------------------------------------------------------------------------------------
\subsubsection{FAPs' Content Placement}
%----------------------------------------------------------------------------------------------------------
Serving requests by FAPs leads to a ground-to-ground connection type between FAPs and ground users. Similarly, the users' access delay through FAP connections is calculated as
\begin{equation} \label{nEq:19}
\mathcal{D}_f(t)=p_k^{(h)}(t) \mathcal{D}_f^{(h)}(t)+p_k^{(m)}(t) \mathcal{D}_f^{(m)}(t),
\end{equation}
where $\mathcal{D}_f^{(h)}(t)$, as the cache-hit delay, is expressed as
\begin{equation} \label{nEq:20}
\mathcal{D}_f^{(h)}(t)= \dfrac{L_c}{R_{i,j}} = L_c \log^{-1}\left (1+\dfrac{P_i |\mathcal{\tilde{H}}_{i,j}(t)|^2 }{I_{\bm{f}_{-i}}(t)+N_0}. \right)
\end{equation}
In this case, coded contents to be stored in the storage of FAPs are determined according to the solution of the following optimization problem:
\begin{eqnarray}{}\label{nEq:21}
\mathcal{F}(\mathbf{y},\mathbf{z}) &\!\!=\!\!&\min \limits_{\substack{y_l, z_l}}~~ \sum \limits_{l=1}^{N_c} \Big ( \sum \limits_{j=1}^{N_g}
p_l^{(j)}(t)  \mathcal{D}_{f}^{(j)} (t)\Big ) y_l  \label{nEq:21}\\
&&+\sum \limits_{l=1}^{N_c} \Big ( \sum \limits_{j=1}^{N_g}
p_l^{(j)}(t) \mathcal{D}_{f}^{(j)} (t)\Big ) z_l,  \nonumber\\ \nonumber
\text{s.t.}~~ &&\textbf{C1.}~~  y_l, z_l \in \{0,1\},
\\ \nonumber
&&\textbf{C2.}~~ \sum \limits_{l=1}^{N_c} (1-y_l) \leq \lfloor \alpha C_f \rfloor, \\ \nonumber
&&\textbf{C3.} ~~\sum \limits_{l=1}^{N_c} (1-z_l) \leq N_s ( C_f - \lfloor \alpha C_f \rfloor),
\end{eqnarray}
where $\mathcal{F}(\mathbf{y},\mathbf{z})$ is the cost function associated with users' access delay, experienced by serving the request through FAPs. By assuming that $N_p=\lfloor \alpha C_f \rfloor $ and  $N_a= N_s ( C_f - \lfloor \alpha C_f \rfloor)$ are the cardinality of popular and mediocre contents, respectively, $\mathbf{y}=[y_1, \ldots, y_{N_p}]^T$ is an indicator vector for popular contents, where $y_l$ would be 0 if $l^{th}$ content is stored in the cache of FAPs, otherwise it equals to~1. Similarly, $\mathbf{z}=[z_1, \ldots, z_{N_a}]^T$ is an indicator variable for mediocre contents. According to the optimization problem, different from popular contents that are completely stored, just one segment of mediocre contents are cached. Similarly, $y_l$ and $z_l$ in constraint \textbf{C1} illustrate the availability of content $c_l$ in the cache of FAP $f_i$. Finally, constraints \textbf{C2} and \textbf{C3} indicate the portion of cache allocated to popular and mediocre contents, respectively. Due to the large size of video contents and the complexity of the content placement, it is essential to update the storage of caching nodes in the off-peak period~\cite{Zhang2021}. Therefore, we use an adaptive time window for cache updating, introduced in our previous work~\cite{Hajiakhondi2019}, to maintain a trade-off between the on-time popularity recognition of contents and the network's traffic.

%=====================================================
\subsection{Content Placement in Multiple Inter-Clusters}
%=====================================================
%%%%%%%%%%%%%%%%%%%%%%%%%%%%%%%%%%%%%%%%%
\begin{figure} [t!]
\centering \includegraphics [scale = 0.27] {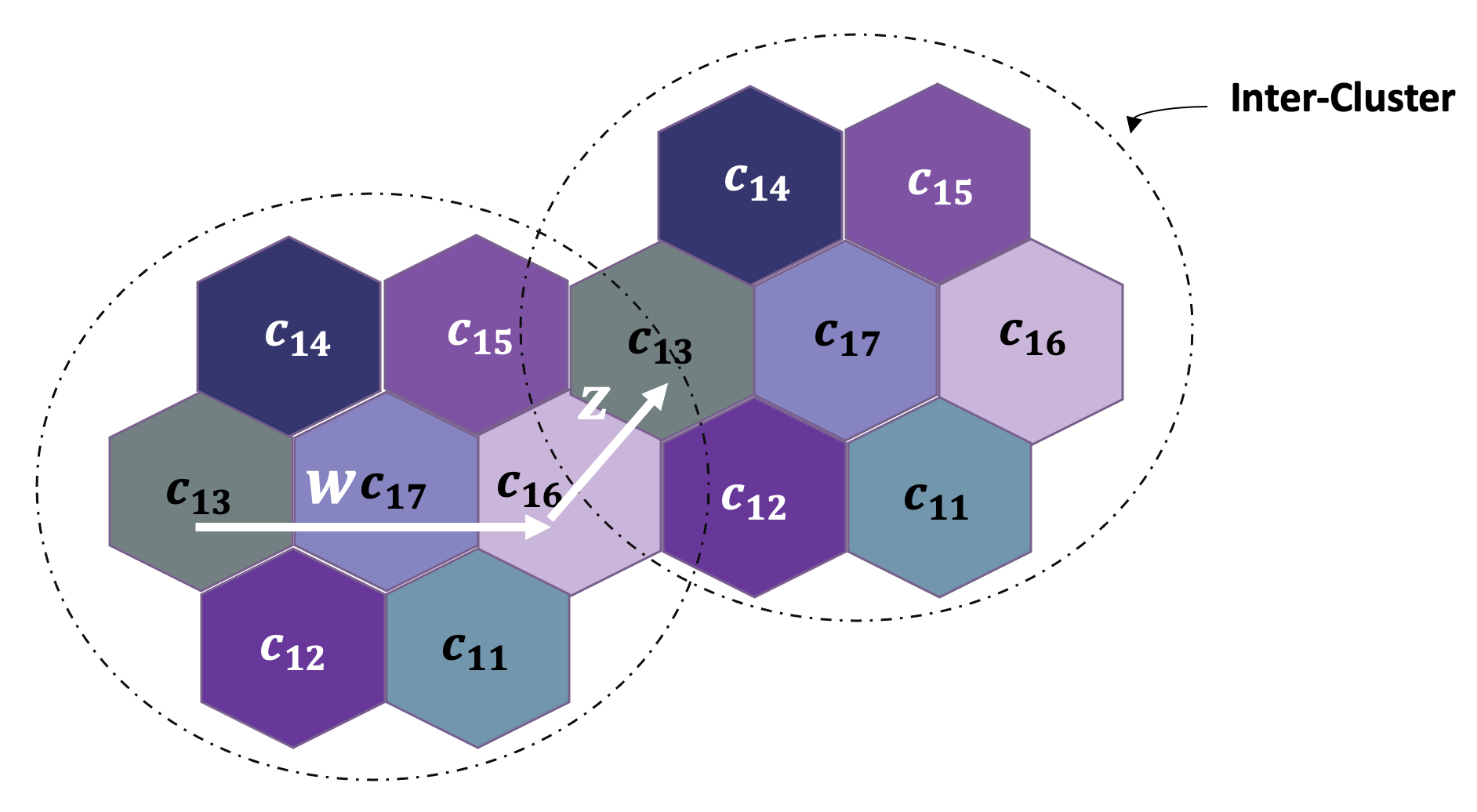}
\vspace{-.2in}
\caption{\footnotesize A typical hexagonal cellular network, where seven FAPs form an inter-cluster.} 
\label{Big_Picture}
\end{figure}
%%%%%%%%%%%%%%%%%%%%%%%%%%%%%%%%%%%%%%%%%
After identifying popular and mediocre contents, we need to determine how to store different segments of mediocre contents within (i) An inter-cluster, and; (ii) Multiple inter-clusters.

\vspace{.05in}
\noindent
\textbf{Single Inter-Cluster}: The main idea behind the coded placement scheme in our proposed CCUF framework comes from the frequency reusing technique in cellular networks~\cite{Choi2006}. The distance between two cells with the same spectrum bandwidth is determined in such a way that the resource availability increases and the inter-cell interference decreases~\cite{Wireless}. With the same argument in~\cite{Choi2006, Wireless}, the same mediocre contents are stored in different inter-clusters, while different FAPs belonging to an inter-cluster store different segments of the mediocre contents.  Without loss of generality, we first consider a simple hexagonal cellular network including $N_b$ FAPs as one inter-cluster (Fig.~\ref{Big_Picture}). Given the vector $\mathbf{z}=[z_1, \ldots, z_{N_a}]^T$ that determines the mediocre contents, in this phase, we need to indicate which segment of the mediocre content $c_l$, denoted by $c_{ls}$ for $(1 \leq l \leq N_a)$ and $(1 \leq s \leq N_s)$, should be cached in FAP $f_i$ for $(1 \leq i \leq N_b)$. In this regard, we form an $(N_a \times N_s)$ indicator matrix of FAP $f_i$, denoted by $\bm{Z}^{(f_i)}$, where the $l^{\text{th}}$ row of $\bm{Z}^{(f_i)}$, denoted by $\bm{z}^{(f_i)}_{l}=[0,0,\ldots, 1]_{(1 \times N_s)}$ indicates segments of file $c_l$ stored in the cache of FAP $f_i$. Note that $\bm{z}^{(f_i)}_{l}$ is a zeros vector with only one non-zero element, where $z^{(f_i)}_{ls} = 1$ means that  $s^{\text{th}}$ segment of file $c_l$ is stored in the cache of FAP $f_i$. To store different segments of mediocre contents within an inter-cluster, the cached contents of FAP $f_j$ for ($1 \leq j \leq N_b, j \neq i$) in the inter-cluster is determined as follows
\begin{equation}\label{nEq:22}
\bm{z}^{(f_i)}_{l} {\bm{z}_{l}^{(f_j)}}^{T}=0,~~~~i=1,\ldots,N_b, j=1,\ldots,N_b, i\neq j.
\end{equation}

\vspace{.05in}
\noindent
\textbf{Multiple Inter-Clusters}: After allocating mediocre contents to FAPs inside an inter-cluster, the same content as FAP $f_i$ is stored in FAP $f_k$ in the neighboring inter-cluster, where $k$ is given by
\begin{equation}\label{nEq:23}
\bm{Z}^{(f_{k})}= \bm{Z}^{(f_i)} ~~~ \text{if}~~~ k=w^2+wz+z^2,
\end{equation}
where $w$ and $z$ represent the number of FAPs required to reach another FAP storing similar contents, in two different directions~\cite{Wireless} (see Fig.~~\ref{Big_Picture}). More precisely, first, it is required to move $w$ cell along any direction from FAP $f_i$, then turn $60$ degrees counter-clockwise and move $z$ cells to reach FAP $f_k$. For example, in Fig.~\ref{Big_Picture}, $N_s = 7$, $w=2$, and $z = 1$ for starting in a FAP including $c_{13}$ and reaching a similar FAP in a neighboring inter-cluster. 

\vspace{.05in}
\noindent
\textbf{Remark 1}: In a practical scenario, the coverage area of FAPs is influenced by path loss and shadowing models (it is not a hexagonal shape). Location $p=(x,y)$ is placed within the transmission area of FAP $f_i$, if the strength of the received signal at point $p$, denoted by $RSSI_{p}$, is higher than the threshold value $RSSI_{th}$, where $RSSI_{p}$ is calculated~as
\begin{equation}
RSSI_{p}(dB)= RSSI(d_0)+10 \eta \log_{10}(\dfrac{d}{d_0})+X_{\sigma},
\end{equation}
with $d$ and $d_0$ denoting the distance between FAP and point $p$ in the boundary of transmission area of FAP, and the reference distance is set to $1$ m, respectively. Moreover, $\eta$ represents the path loss exponent, which is $10$ dB or $20$ dB, and $X_{\sigma}$ is a zero-mean Gaussian with standard deviation $\sigma$ that represents the effect of multi-path fading in the CCUF scheme~\cite{Al-Samman2019}.

%=====================================================
\subsection{Success Probability in the Proposed CCUF Framework}
%=====================================================
To quantify the benefits of the proposed CCUF strategy, we define success probability, which is defined as the probability of finding a new segment by user $GU_j$ at time slot $t$ under the following two scenarios: (i)  Uncoded cluster-centric, and; (ii)  Coded cluster-centric UAV-aided femtocaching network, denoted by $p_{uc}$, and $p_{cc}$, respectively. Concerning the nature of mobile networks, ground users move and leave their current positions. In this paper, it is assumed that low-speed ground users can obtain one segment in each contact, i.e.,  $T=N_s\delta_t$ is required to completely download content $c_l$. First, we consider a simple mobility, where ground users are positioned in the transmission area of a new FAP in each time slot $t$. Eventually, in $T=N_s\delta_t$, the whole content $c_l$ will be downloaded. Then, we generalize the mobility of ground users to the random walk model, where ground users can return to their previous place.

%----------------------------------------------------------------------------------------------------------
\subsubsection{Simple Movement Scenario}
%----------------------------------------------------------------------------------------------------------
Regarding the uncoded cluster-centric UAV aided femtocaching framework, content $c_l$ consisting of $c_{ls}$, for ($1 \leq s \leq N_s$) segments, is stored completely in all FAPs. Consequently, the probability of downloading $n= N_s$ segments of content $c_l$ in $T=N_s\delta_t$ depends on the probability of requesting file $c_l$, denoted by $p_l$. Since the storage capacity of each FAP is equal to $C_{f}$, the success probability, denoted by $p_{uc}$, is obtained as follows
\begin{equation}\label{nEq:25}
p_{uc}[n=N_s , t= T]=\sum \limits_{l=1}^{C_f}p_l.
\end{equation}
On the other hand, the success probability of the CCUF framework is obtained as
\begin{equation}\label{nEq:26}
p_{cc}[n=N_s , t= T]=\sum \limits_{l=1}^{N_p}p_l+\sum \limits_{l=N_p+1}^{N_a+N_p}p_l.
\end{equation}
To illustrate the growth rate of the success probability in the coded one, we rewrite $p_{uc}$ in Eq.~\eqref{nEq:25} as follows
\begin{equation}\label{nEq:27}
p_{uc}[n=N_s , t= T]=\sum \limits_{l=1}^{N_p}p_l+\sum \limits_{l=N_p+1}^{C_f}p_l.
\end{equation}
As it can be seen from Eqs.~\eqref{nEq:26}, and~\eqref{nEq:27}, the first term related to the popular content is the same. The second term, however, illustrates that the number of distinct contents that can be served through FAPs within an inter-cluster in the coded cluster-centric network is $\varkappa$ times greater than the uncoded one, where $\varkappa$ is given by
\begin{equation}\label{nEq:28}
\varkappa=\dfrac{\lfloor \alpha C_f \rfloor+ N_s ( C_f - \lfloor \alpha C_f \rfloor)}{C_f}.
\end{equation}
Accordingly, due to the allocation of different segments in the coded cluster-centric network, more segments of the desired contents are accessible during the users' movement in the simple movement scenario. Therefore, more requests can be served in comparison to the uncoded cluster-centric UAV-aided femtocaching networks.

%----------------------------------------------------------------------------------------------------------
\subsubsection{Generalizing to Random Walk Scenario}
%----------------------------------------------------------------------------------------------------------
In contrary to the simple movement scenario discussed above, the following two  situations are possible where the ground user $GU_j$ cannot find a new segment during its movement: (i) Returning back to the previous coverage area of FAPs; and, (ii) Positioning in the transmission area of a FAP, which stores the same segment of the content that the ground user has already downloaded. Consequently, the success probability of the coded cluster-centric will not be the same as the previous scenario. If the requested content is the popular one, regardless of the link's quality of the ground user within an inter-cluster, the ground user can download one segment of the required content with the probability of $\sum \limits_{l=1}^{N_p}p_l$ at each contact. While this part of the success probability is constant,  the success probability of downloading a new segment of a mediocre content in each contact depends on the current and previous locations of the ground user. Therefore, we first determine the success probability of achieving a new segment of a mediocre content, denoted by $p_{ns}(n=n_0, t=n_0 \delta_t)$, for ($1 \leq n_0 \leq N_s)$. Then, we calculate the success probability of a coded cluster-centric network based on the random walk movement.

As it can be seen from Fig.~\ref{Big_Picture}, regardless of the location of $GU_j$, this user can download one segment successfully in the first contact (i.e., $n_0=1$). Therefore, we have $p_{ns}(n=1, t=\delta_t)=1$. Similarly, when $n_0=2$, the ground user $GU_j$ can download a new segment without considering the location of the ground user. Therefore, the probability of downloading two segments after two contacts is $p_{ns}[n=2 , t= 2\delta_t]=1$. More precisely, in the second contact, the ground user can be positioned in the cell of $(N_s-1)$ number of FAPs, where the probability of being in the cell of FAP $f_i$ is $p(f=f_i)= \dfrac{1}{(N_s-1)}$. Therefore, we have
\begin{eqnarray}\label{nEq:29}
p_{ns}[n=2, t=2\delta_t] &\!\!\!\!\!=\!\!\!\!\!\!\!\!&\sum_{i=1}^{N_s-1} \!\!p_{ns}[n=2, t=2\delta_t|f=f_i] p(f=f_i) \nonumber\\
&&\!\!\!\!\!\!\!\!\!\!\!\!\!\!\!\!\!\!\!\!\!\!\!\!\!\!\!\!\!\!\!\!\!\!\!\!  (N_s-1)\times 1\times \dfrac{1}{N_s-1}+ 0\times \dfrac{1}{N_s-1}=1.
\end{eqnarray}
Accordingly, the probability of finding a new segment in the third contact is obtained as follows
\begin{eqnarray} \label{nEq:30}
\lefteqn{\!\!\!\!\!\!\!\!\!\!\!p_{ns}[n=3, t=3\delta_t]\!=\!\!\!\!\sum_{i=1}^{N_s-1}\!\! p_{ns}[n=3, t=3\delta_t|f=f_i] p(f=f_i) \nonumber}\\
&&\!\!\!\!\!\!\!\!\!\!\!\!\!\!\!\!\!\!\!\!\!=(N_s-2) \dfrac{1}{N_s-1}\times1+\dfrac{1}{N_s-1}\times0= \dfrac{N_s-2}{N_s-1},
\end{eqnarray}
where $(N_s-2)$ FAPs have different segments, whereas if $GU_j$ returns to the FAP at $t=\delta_t$, the ground user can find a similar segment. Similarly, it can be proved that the probability of finding a new segment in $n>2$ is given by
\begin{eqnarray}\label{nEq:31}
\lefteqn{\!\!\!\!\!\!\!\!\!\!\!\!\!\!\!\!\!\!\!\!\!\!p_{ns}[n=n_0, t=n_0 \delta_t]= \sum_{i=1}^{N_s-1} p_{ns}[n=n_0, t=n_0 \delta_t|f=f_i]  \nonumber}\\
&&\!\!\!\!\!\!\!\!\!\!\!\!\!\!\!\!\!\!\!\!\! \times p(f=f_i) = \dfrac{(N_s-2)^{n_0-2}}{(N_s-1)^{n_0-2}},~~~~~\text{for}~n_0>2.
\end{eqnarray}
Taking into account the unequal likelihood of finding new segments of mediocre contents in different contacts, we recalculate $p_{cc}$ as follows
\begin{eqnarray}\label{nEq:32}
\lefteqn{p_{cc}[n=N_s , t= T] =\sum \limits_{l=1}^{N_p}p_l \nonumber }\\
&+&\sum \limits_{n=1}^{N_s} \dfrac{(N_s-2)^{n-2}}{(N_s-1)^{n-2}} \Big (\sum \limits_{l=N_p+1}^{N_s ( C_f - \lfloor \alpha C_f \rfloor)}p_l \Big ).
\end{eqnarray}
%
%\vspace{.05in}
%\noindent
\textbf{Remark 2}: In such a case that the location of $GU_j$ in two consecutive time slots is the same, it means that $GU_j$ is a fixed user. Therefore, the following two scenarios can happen depending on the popularity of the requested content: (i) Similar to the mobile users' case, if the requested content $c_l$ is popular, the whole segments of file $c_l$ are sent by neighboring FAP to $GU_j$, and; (ii) If $c_l$ is a mediocre content, each FAP within the inter-cluster transmits one segment of file $c_l$, which is known as the Parallel Transmission (PT)~\cite{Chen2017_2}.
%%%%%%%%%%%%%%%%%%%%%%%%%%%%%%%%%%%%%%%%%
\setlength{\textfloatsep}{0pt}
\begin{algorithm}[t!]
\caption{Proposed CCUF Strategy}
\begin{algorithmic}[1]
\State \textbf{Initialization:} Set $\alpha$, $\lambda$, $N_s$, and $C_f$.
\State \textbf{Input:} $p_l^{(j)}(t)$. %, $\bm{L}_{k}^{(t)}$, and $\bm{L}_{j}^{(t)}$.
\State \textbf{Output:} $\mathrm{x}_{l}$, $\mathtt{y}_{l}$, and $\mathtt{z}_{l}$.
\State \textbf{Content Placement Phase:}
\For {$ u_k, k=1,\ldots,N_u $,}
 $$\min \limits_{\substack{\mathrm{x}_l}}~~ \sum \limits_{l=1}^{N_c} \Big ( \sum \limits_{j=1}^{N_g}
p_l^{(j)}(t)  \mathcal{D}_{u}^{(j)} (t)\Big ) \mathrm{x}_l $$\\
~~~~~~~~~~\text{s.t.}~~ \textbf{C1.} and \textbf{C2.} in Eq.~\eqref{nEq:18}.
\EndFor
\For {$ f_i, i=1,\ldots,N_f $,}
$$\min \limits_{\substack{y_l, z_l}}~~ \sum \limits_{l=1}^{N_c} \Big ( \sum \limits_{j=1}^{N_g}
 p_l^{(j)}(t)  \mathcal{D}_{f}^{(j)} (t)\Big ) y_l +$$
$$\sum \limits_{l=1}^{N_c} \Big ( \sum \limits_{j=1}^{N_g}
p_l^{(j)}(t)  \mathcal{D}_{f}^{(j)} (t)\Big ) z_l,$$\\
~~~~~~~~~~\text{s.t.}~~ \textbf{C1.-C3.}~~in Eq.~\eqref{nEq:21}.
\EndFor \\
$\bm{z}^{(f_i)}_{l} {\bm{z}_{l}^{(f_j)}}^{T}=0,~~~~i=1,\ldots,N_s, j=1,\ldots,N_s, i\neq j,$ \\
$\bm{Z}^{(f_{k})}= \bm{Z}^{(f_i)} ~~~ \text{if}~~~ k=w^2+wz+z^2,$
\State \textbf{Transmission Phase:}
\For {$ GU_j, j=1,\ldots,N_g $,}
\If{$GU_j$ is in indoor environment}
\If{$GU_j$ is an edge-user and requests \\
~~~~~~~~~popular content}
\State {The request should be handled according to the\\
~~~~~~~~~~~~~JT scheme.}
\Else
\State {The request should be handled according to the\\
~~~~~~~~~~~~~ST scheme.}
\EndIf
\Else
\If{$\upsilon_{j}(t) \geq \upsilon_{\text{th}}$} \\
\State {The request is served by UAV $u_k$.}
\Else
~~~~~~~~\State{Similar to lines $16$ to $23$.}
\EndIf
\EndIf
\EndFor
\end{algorithmic}
\label{algo1}
\end{algorithm}
\vspace{-.1in}
%%%%%%%%%%%%%%%%%%%%%%%%%%%%%%%%%%%%%%%%%

%=====================================================
\subsection{2-D Deployment of UAVs in Intra-clusters}
%=====================================================
To increase the resource availability for ground users, the outdoor environment is partitioned based on an unsupervised learning algorithm, where each partition is covered by a UAV. Considering a Gaussian mixture distribution for ground users, we have  a dense population of ground users in some areas. The main goal is to deploy UAVs in such a way that ground users can experience high QoS communications even in a dense area. Note that the distance between UAVs and ground users is a critical factor that can significantly impact the QoS from different perspectives such as the energy consumption of UAVs and the users' access delay. Our goal is to partition $N_g$ ground users into $K$ intra-clusters, where the sum of Euclidean distances between the ground user $GU_j$, for $(1 \leq j \leq N_g^{k})$, and UAV $u_k$ is minimized. In this case, $N_g^{k}$ is the cardinality of ground users positioned in the intra-cluster related to the UAV $u_k$. Therefore, the UAVs' deployment is obtained according to the following optimization problem
\begin{eqnarray}\label{nEq:33}
\min \limits_{\substack{\bm{l}_{k}(t)}}~~ \sum \limits_{k=1}^{N_u} \sum \limits_{j=1}^{N_g^k} || \bm{l}_{j}(t) , \bm{l}_{k}(t) ||,
\end{eqnarray}
where $\bm{l}_{k}(t)$ denotes the location of the UAV $u_k$ at time slot $t$, defined as the mean of the coordinates of all ground users inside the corresponding intra-cluster as follows
\begin{eqnarray}\label{nEq:34}
\bm{l}_{k}(t)= \dfrac{\sum \limits_{j=1}^{N_g^k} \bm{l}_{j}(t)}{N_g^k}, ~~~~~~ k=1,\ldots,N_u.
\end{eqnarray}
To solve the above optimization problem, we utilize the $K$-Means clustering algorithm~\cite{Kanungo2002}, which is known as an efficient unsupervised learning framework. In the first step, a set of points, denoted by $\mathcal{P}=\{P_1,\ldots,P_{N_u}\}$, is generated, where $P_k$ for ($1 \leq k \leq N_u$) should be within the pre-specified environment. Then, the set of ground users in the vicinity of $P_k$ is determined as follows
\begin{eqnarray}\label{nEq:35}
u_j \in N_g^k ~~~~ \text{if} ~~||\bm{l}_{j}(t) , P_k || < ||\bm{l}_{j}(t) , P_r || ,~~	\forall k \neq r.
\end{eqnarray}
Given the set of ground users belonging to each intra-cluster, UAVs' locations are determined according to Eq.~\eqref{nEq:34}. In the second step, by moving ground users from one intra-cluster to another, the Euclidean distances between ground users and UAVs are calculated to update the location of UAVs according to Eq.~\eqref{nEq:33}. The $K$-Means algorithm is terminated when  there is no change in the ground users belonging to an intra-cluster over several iterations. This completes our discussion on development of the CCUF scheme. The pseudo-code of the proposed CCUF framework is summarized in \textbf{Algorithm 1}.

%OOOOOOOOOOOOOOOOOOOOOOOOOOOOOOOOOOOOOOOOOOOOOOOOOOOOOOOOO
\section{Simulation Results} \label{sec:sim}
\begin{figure} [t!]
\centering \includegraphics [scale = 0.3] {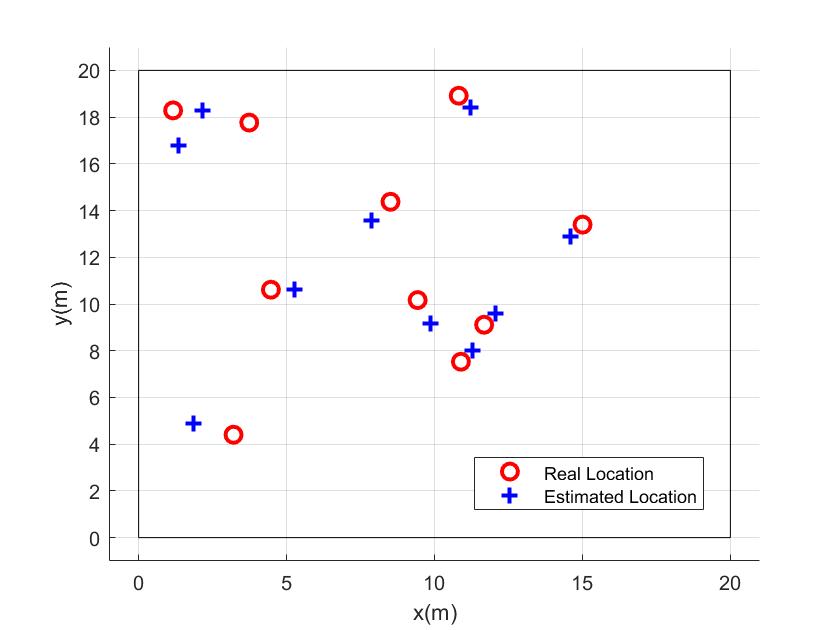}
\caption{\footnotesize Typical location estimation results based on the AoA localization scheme.} \label{FL}
\end{figure}

%OOOOOOOOOOOOOOOOOOOOOOOOOOOOOOOOOOOOOOOOOOOOOOOOOOOOOOOOO
To demonstrate the advantage of the proposed CCUF framework, we 
consider a UAV-aided cellular network with $R=5000$ m, covered by the main server. There are $N_f=175$ FAPs, and $N_u=10$ UAVs, where each inter-cluster compromises of $N_s=7$ FAPs. Without loss of generality and for simplicity, we consider a static clustering scheme where there are a fixed number of FAPs in each inter-cluster to determine how different segments of popular files should be distributed in an inter-cluster to increase the content diversity. Therefore,  the number of FAPs in each inter-cluster is considered to be the same as the number of segments $N_s$. According to the  restrictions of the  aviation regulations, UAVs fly horizontally at the height of  $h_k= 100$ m, covering a region of the outdoor environment with $R_u= 500$ m, while the transmission range of FAPs is $30$ m~\cite{Hajiakhondi2021}. The general simulation parameters are summarized in Table~\ref{table1}. As it is proved in~\cite{Shanmugam2013, Hajiakhondi2019} that the optimum content placement is an NP-hard problem, we use \textit{fmincon} optimization toolbox, implemented in MATLAB (R2020a), to solve Eqs.~\eqref{nEq:18} and~\eqref{nEq:21}.  Fig.~\ref{FL} illustrates a typical $20 \times 20$ $m^{2}$ area, where ground users are randomly distributed and move according to~\cite{Albertsen2019}. The estimated location of GUs is required to determine the transmission scheme and select an appropriate caching node to manage the request. Capitalizing on the reliability and efficiency of the AoA localization technique~\cite{Hajiakhondi2020_1, Hajiakhondi2020_2}, it is utilized to estimate the GUs' locations. It can be shown that the Root Mean Square Error (RMSE) between the estimated and the actual location of the ground users is about $0.4$ m, which is acceptable in comparison to the transmission range of FAPs. 

%%%%%%%%%%%%%%%%%%%%%%%%%%%%%%%%%%%%%%%%%
%%%%%%%%%%%%%%%%%%%%%%%%%%%%%%%%%%%%%%%%%

\begin{figure} [!t]
\centering \includegraphics [scale = 0.4] {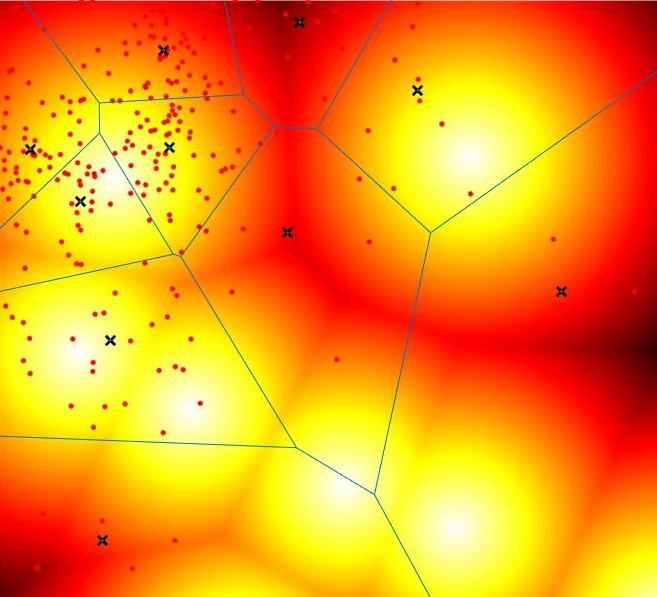}
\caption{\footnotesize Deployment of UAVs in intra-clusters within an integrated network, where ``yellow'' and ``red'' colors indicate indoor and outdoor environments, respectively.} \label{deployment}
\end{figure}

Fig.~\ref{deployment} depicts an integrated heterogeneous network, where yellow and red areas determine indoor and outdoor environments, respectively. Fig.~\ref{deployment} also shows the deployment of UAVs in the intra-clusters within the network, which is generated by partitioning ground users according to the K-means clustering algorithm. As a result of the Gaussian mixture distribution for clients, we have a dense population in some areas, which can be changed over time by the movement of ground users. Therefore, the location of $N_u=10$ UAVs and the formation of intra-clusters in this paper is varying, depending on the user density distribution. For the comparison purpose and in order to find the best value of $\alpha$, three types of caching strategies are considered:
\begin{itemize}
\item \textit{Uncoded UAV-aided Femtocaching (UUF):} This scheme is derived by modifying the Fairness Scheduling algorithm with an Adaptive Time Window (FS-ATW) scheme~\cite{Hajiakhondi2019}, where the proposed content placement strategy in~\cite{Hajiakhondi2019} is used for both UAVs and FAPs. The popular contents in the UUF model are stored completely into FAPs and UAVs without any coding and clustering schemes. Therefore, it is equivalent to our proposed CCUF framework, where the value of $\alpha$, which indicates the percentage of contents stored completely, would be one (i.e., $\alpha=1$).
\item \textit{Proposed Cluster-centric and Coded UAV-aided Femtocaching (CCUF):} In this case, the uncoded popular and the coded mediocre contents are stored in the caching nodes, where $0 <\alpha < 1$. According to the simulation results, the best value of $\alpha$ is obtained.
\item \textit{The Conventional Cluster-centric and Coded UAV-aided Femtocaching (Conventional CCUF):} This scheme is an upgraded version of the FemtoCaching scheme in~\cite{Shanmugam2013}, integrated with the CoMP technology. In this framework, regardless of the content popularity profile, all contents are stored partially. For simplicity, this scheme is shown by $\alpha=0$ in simulation results.
\end{itemize}
These three strategies are evaluated over the cache-hit-ratio, cache diversity, cache redundancy, SINR, and users' access delay to determine the best value of $\alpha$. Moreover, to illustrate the effect of considering a UAV-aided femtocaching framework in an integrated network, we compare the users' access delay and energy consumption of UAVs, by serving users in both indoor and outdoor areas.

%%%%%%%%%%%%%%%%%%%%%%%%%%%%%%%%%%%%%%%%%
%%%%%%%%%%%%%%%%%%%%%%%%%%%%%%%%%%%%%%%%%

%%%%%%%%%%%%%%%%%%%%%%%%%%%%%%%%%%%%%%%%%
\begin{table}[t!]
\caption{List of Parameters.} \label{table1}
\centering
\begin{tabular}{|c|c|c|c|}
\hline
\textbf{Notation} & \textbf{Value} & \textbf{Notation} & \textbf{Value}\\
\hline
$N_g$ & $500$ & $\eta^{(LoS)}$, $\eta^{(NLoS)}$  & $2.5$, $3$ \\ \hline
$N_f$ & $180$ & $h_k$ & $100$ m\\ \hline
$N_u$ & $10$ & $\varpi$, $\psi$ & $2$, $20$ \\ \hline
$N_s$ & $7$  & $L_c$ & $37.5$ MB\\ \hline
$N_c$ & $40724$ &  $\tau_p$ & $0-5$ s \\ \hline
$R_u$ & $500$ m & $P_k$ & $15$ dBm\\ \hline
$R_f$ & $30$ m  & $\chi_{\sigma}^{(LoS)}$, $\chi_{\sigma}^{(NLoS)}$ & $3.5$, $3 $\\ \hline
$P_{T}(t)$, $P_{R}(t)$ & $0.5$ , $0.25$ W & $N_0$ & $-94$ dBm  \\ \hline
\end{tabular}
\end{table}
%%%%%%%%%%%%%%%%%%%%%%%%%%%%%%%%%%%%%%%%%

%%%%%%%%%%%%%%%%%%%%%%%%%%%%%%%%%%%%%%%%%
\begin{figure} [!t]
\centering \includegraphics [scale = 0.6] {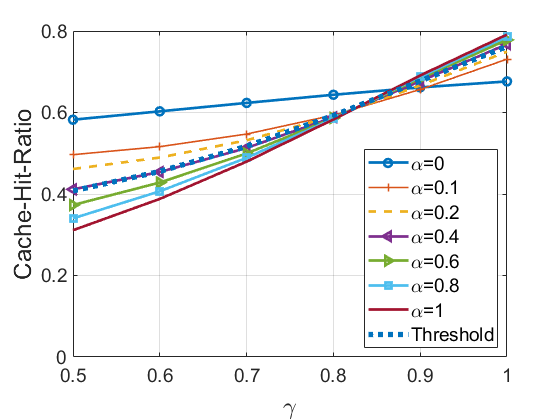}
\caption{\footnotesize The cache-hit-ratio versus the popularity parameter $\gamma$ for different values of $\alpha$.} \label{CHR1}
\vspace{-.1in}
\end{figure}
%%%%%%%%%%%%%%%%%%%%%%%%%%%%%%%%%%%%%%%%%

%%%%%%%%%%%%%%%%%%%%%%%%%%%%%%%%%%%%%%%%%
\begin{figure} [t!]
\centering \includegraphics [scale = 0.6] {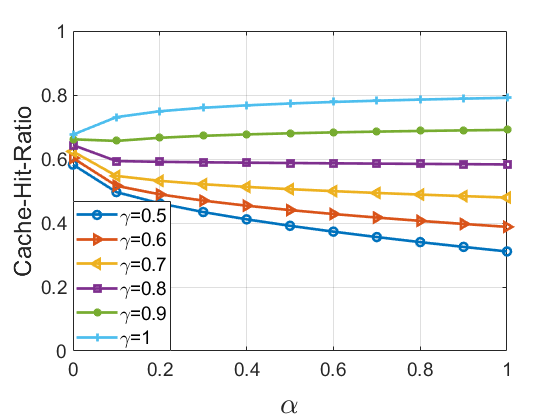}
\caption{\footnotesize The cache-hit-ratio versus the $\alpha$ percentage of contents that are stored completely.} \label{CHR2}
\end{figure}
%%%%%%%%%%%%%%%%%%%%%%%%%%%%%%%%%%%%%%%%%

\vspace{.05in}
\noindent
\textit{Cache-Hit-Ratio:} This metric illustrates the number of requests served by caching nodes versus the total number of requests made across the network. The high value of cache-hit-ratio shows the superiority of the framework. Since we assume that ground users  can  download one segment in each contact, we evaluate the cache-hit-ratio in terms of the number of fragmented contents served by caching nodes. Fig.~\ref{CHR1} compares the cache-hit-ratio of the UUF ($\alpha=1$), the proposed CCUF ($0 <\alpha < 1$), and conventional CCUF ($\alpha=0$) frameworks versus the value of $\gamma$. As previously mentioned, parameter  $\gamma$ shows the skewness of the content popularity, where $\gamma \in [0,1]$. Note that the large value of $\gamma$ indicates that a small number of contents has a high popularity, where a small value of $\gamma$ illustrates an almost uniform popularity distribution for the majority of contents. As it can be seen from Fig.~\ref{CHR1}, depending on the popularity distribution of contents, $\gamma$, the conventional CCUF framework results in a higher cache-hit-ratio. The most important reason is that given a constant cache capacity, the coded content placement of the conventional CCUF strategy leads to a remarkable surge in the content diversity. In contrast, for a high value of $\gamma$, where a small number of contents is widely requested, the UUF and the proposed CCUF frameworks have better results compared to the conventional CCUF. By considering the fact that the common value of $\gamma$ is about $0.5  \leq \gamma \leq 0.6$ (e.g., see~\cite{Hajiakhondi2020, Shanmugam2013}), we define $CHR_{th}$ as the threshold cache-hit-ratio, which is the average of cache-hit-ratio of different values of $\alpha$ for a specific $\gamma$. As it can be seen from Fig.~\ref{CHR1}, the proposed CCUF framework with $0 < \alpha \leq 0.4$  and the UUF scheme outperform other schemes from the aspect of cache-hit-ratio.

Fig.~\ref{CHR2} shows the cache-hit-ratio versus different values of $\alpha$ when the popularity parameter $\gamma$ changes in the range of $0.5$ to $1$. Accordingly, for $0.5  \leq \gamma \leq 0.6$, by increasing the value of $\alpha$, the cache-hit-ratio decreases drastically. In the following, we also investigate the impact of $\alpha$ on the users' access delay to determine the best value of $\alpha$.

%%%%%%%%%%%%%%%%%%%%%%%%%%%%%%%%%%%%%%%%%%
\begin{figure} [!t]
\centering \includegraphics [scale = 0.6] {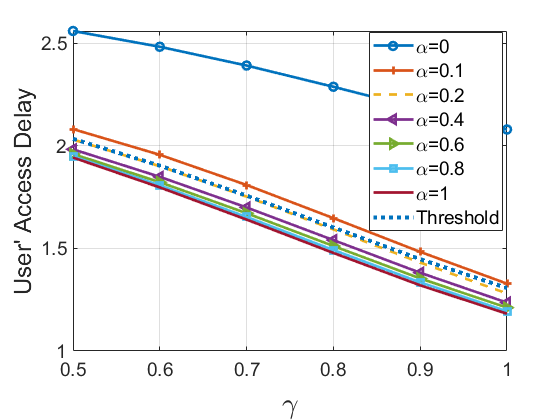}
\caption{\footnotesize The users' access delay in the indoor environment versus different value of $\gamma$.} \label{Delay2}
\vspace{-.1in}
\end{figure}
%%%%%%%%%%%%%%%%%%%%%%%%%%%%%%%%%%%%%%%%%%

%%%%%%%%%%%%%%%%%%%%%%%%%%%%%%%%%%%%%%%%%%
\begin{figure} [t]
\centering \includegraphics [scale = 0.6] {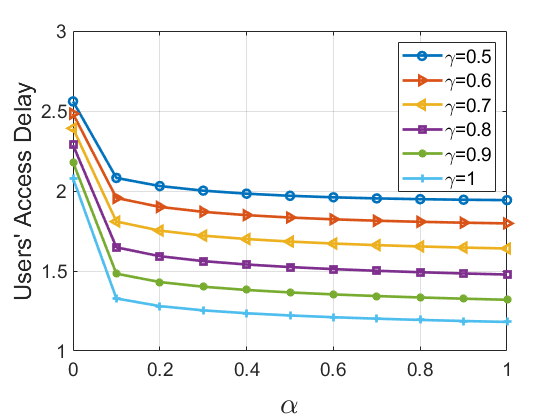}
\vspace{-.2in}
\caption{\footnotesize The users' access delay in the indoor environment versus different values of $\alpha$.} \label{Delay1}
\end{figure}
%%%%%%%%%%%%%%%%%%%%%%%%%%%%%%%%%%%%%%%%%%

\noindent
\textit{Users' Access Delay:} Users' access delay depends on three parameters, i.e., the availability of the content in caching nodes, the distance between the ground user and the corresponding caching node, and the channel quality, known as the SINR. Figs.~\ref{Delay2} and~\ref{Delay1} compare the users' access delay of the aforementioned frameworks, which is obtained according to Eq.~\eqref{nEq:19}. By utilizing the CoMP technology in the proposed CCUF, serving edge-users according to the JT scheme has a great impact on the SINR, where users' access delay decrease by increasing the SINR. As can be seen from Table~\ref{table2}, the SINR of edge-users improves by increasing the value of $\alpha$. Note that JT scheme can be performed if the same contents are stored in the neighboring FAPs. Therefore, by increasing the value of $\alpha$, the users' access delay will decrease. With the same argument, we define $\mathcal{D}_{th}$, which is the average of users' access delay of different values of $\alpha$ for a specific $\gamma$, shown in Fig.~\ref{Delay2}. Therefore, the best value of $\alpha$ would be $\alpha \geq 0.2$. Consequently, the cache-hit-ratio and users' access delay of the proposed CCUF framework would be efficient if $\alpha \in [0.2,0.4]$.

%%%%%%%%%%%%%%%%%%%%%%%%%%%%%%%%%%%%%%%%%%
\begin{table}[!t]
\caption{\footnotesize The SINR experienced by edge-users for different values of $\alpha$ and $\gamma$.} \label{table2}
\centering
\begin{tabular}{|c|l|c|l|c|l|c|l|c|l|c|l|c|l|c|l|}
\hline
& \textbf{$\gamma=0.5$} & \textbf{$\gamma=0.6$} & \textbf{$\gamma=0.7$}& \textbf{$\gamma=0.8$}& \textbf{$\gamma=0.9$}& \textbf{$\gamma=1$}\\
\hline
\textbf{$\alpha=0$} & $16.37$ & $16.37$ & $16.37$ & $16.37$ & $16.37$ & $16.37$ \\
\hline
\textbf{$\alpha=0.1$} & $17.55$& $18.12$ & $18.89$ & $19.84$ & $20.88$ & $21.88$ \\
\hline
\textbf{$\alpha=0.2$}& $18.01$& $18.65$ & $19.46$ & $20.40$ & $21.38$ & $22.30$ \\
\hline
\textbf{$\alpha=0.4$}& $18.62$& $19.30$ & $20.11$ & $21.00$ & $21.90$ & $22.70$ \\
\hline
\textbf{$\alpha=0.6$}& $19.06$& $19.75$ & $20.53$ & $21.37$ & $22.20$ & $22.93$ \\
\hline
\textbf{$\alpha=0.8$} & $19.42$& $20.09$ & $20.85$ & $21.64$ & $22.42$ & $23.08$ \\
\hline
\textbf{$\alpha=1$} & $19.72$& $20.38$ & $21.11$ & $21.86$ & $22.58$ & $23.20$ \\ \hline

\end{tabular}
\end{table}
%%%%%%%%%%%%%%%%%%%%%%%%%%%%%%%%%%%%%%%%%%

%%%%%%%%%%%%%%%%%%%%%%%%%%%%%%%%%%%%%%%%%%
\begin{figure} [!t]
\centering \includegraphics [scale = 0.6] {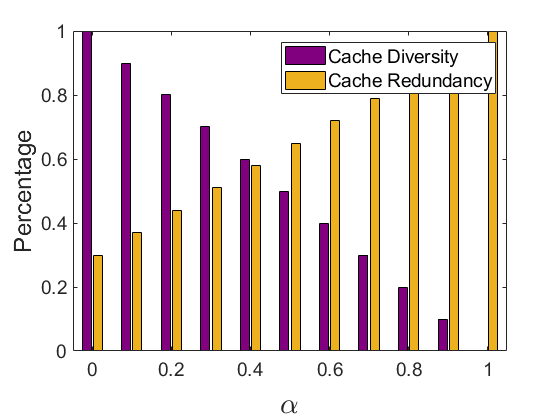}
\caption{\footnotesize The percentage of the cache diversity and the cache redundancy versus different values of $\alpha$.} \label{diversity}
\end{figure}
%%%%%%%%%%%%%%%%%%%%%%%%%%%%%%%%%%%%%%%%%%

%%%%%%%%%%%%%%%%%%%%%%%%%%%%%%%%%%%%%%%%%%
\begin{figure} [t]
\centering \includegraphics [scale = 0.6] {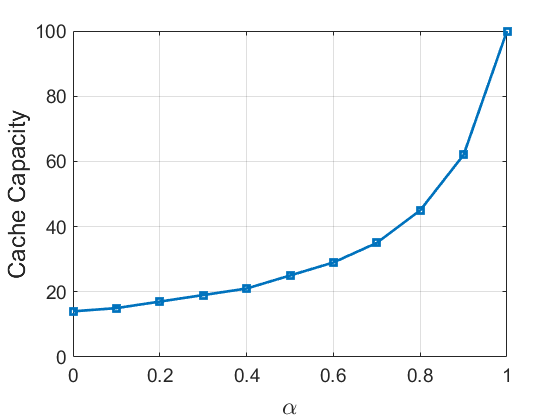}
\caption{\footnotesize The maximum cache capacity, required to achieve the maximum cache diversity, versus different values of $\alpha$.} \label{cache_capacity}
\end{figure}
%%%%%%%%%%%%%%%%%%%%%%%%%%%%%%%%%%%%%%%%%%

\vspace{.05in}
\noindent
\textit{Cache Diversity:} This metric illustrates the diversity of contents in an inter-cluster, which is defined as the number of distinct segments of contents, expressed as follows
\begin{eqnarray}\label{nEq:35}
\mathcal{CD}=\dfrac{N_a}{N_s C_f}=1-\dfrac{\lfloor \alpha C_f \rfloor}{C_f}.
\end{eqnarray}
As stated previously, we have $N_a= N_s ( C_f - \lfloor \alpha C_f \rfloor)$. As it can be seen from Fig.~\ref{diversity}, the value of $\mathcal{CD}$ would be one, if $\alpha=0$, which means that all cached contents are different. The cache diversity, however, linearly decreases by increasing the value of $\alpha$, and reaches the lowest value zero, when all contents are cached completely (i.e., $\alpha=1$).

\vspace{.05in}
\noindent
\textit{Cache Redundancy:} This metric indicates the number of similar contents that ground users meet during their random movements. As it can be seen from Fig.~\ref{diversity}, the cache redundancy increases by storing the entire contents. By considering the coded content placement, even in the proposed CCUF framework, ground users that move randomly through the network, can meet a similar coded contents during their movements (see Fig.~\ref{Big_Picture}).

\vspace{.05in}
\noindent
\textit{Maximum Required Cache Capacity:} Given a specific number of contents through the network, denoted by $N_c$, the storage capacity of caching nodes is determined by $C_f=\beta N_c$. In this case, parameter $\beta$ indicates the percentage of contents that can be stored in caching nodes. In the coded content placement, since  only one segment of the contents is cached, it is fairly likely that the total number of possible segments that can be cached exceeds the total number of contents. Therefore, the maximum required cache capacity, denoted by $\beta_{max}$, for different values of $\alpha$ is obtained as
\begin{equation}
\beta_{max} \leq \dfrac{N_c}{N_s N_c -(N_s-1)\alpha N_c}=\dfrac{1}{\alpha(1-N_s)+N_s},
\end{equation}
where the remainder of the storage would be occupied by redundant contents if $\beta > \beta_{max}$. As it can be seen from Fig.~\ref{cache_capacity}, the maximum cache capacity $\beta_{max}$ increases by the value of $\alpha$. Consequently, in smaller values of $\alpha$, we need a smaller cache capacity to have the maximum cache diversity.
%%%%%%%%%%%%%%%%%%%%%%%%%%%%%%%%%%%%%%%%%%
\begin{figure} [!t]
\centering \includegraphics [scale = 0.6] {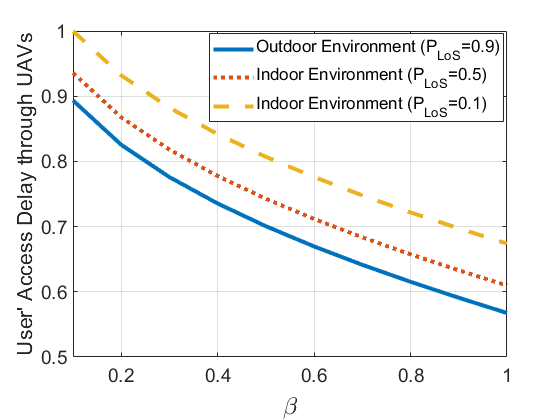}
\caption{\footnotesize The users' access delay experienced through UAVs in both indoor and outdoor versus different values of $\beta$.} \label{Delay_UAV}
%\vspace{-.1in}
\end{figure}
%%%%%%%%%%%%%%%%%%%%%%%%%%%%%%%%%%%%%%%%%%

%%%%%%%%%%%%%%%%%%%%%%%%%%%%%%%%%%%%%%%%%%
\begin{figure} [t]
\centering \includegraphics [scale = 0.6] {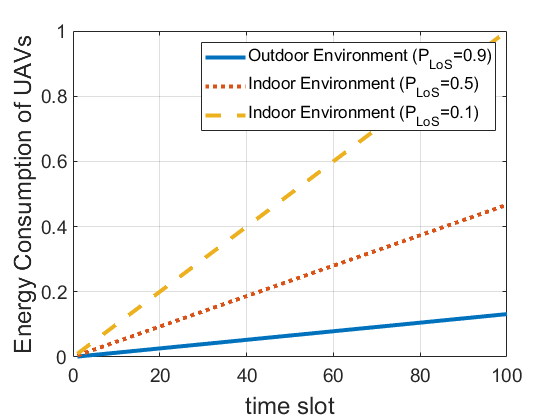}
\caption{\footnotesize The energy consumption of UAVs in both indoor and outdoor environments in different time slots.} \label{energy}
\end{figure}
%%%%%%%%%%%%%%%%%%%%%%%%%%%%%%%%%%%%%%%%%%
%%%%%%%%%%%%%%%%%%%%%%%%%%%%%%%%%%%%%%%%%%
\begin{figure} [t]
\centering \includegraphics [scale = 0.6] {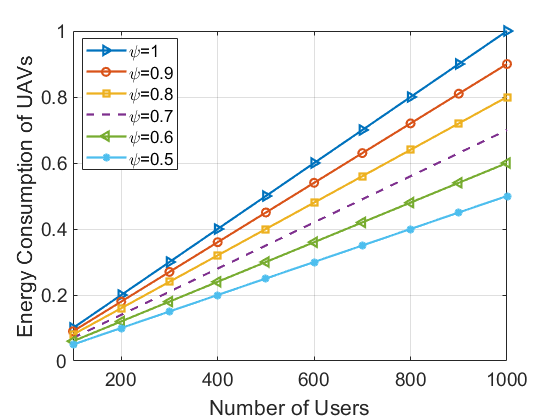}
\caption{\footnotesize The normalized energy consumption of UAVs versus the number of outdoor users, where $\psi$ illustrates the ratio of requests served by UAVs to the whole outdoor users' requests.} \label{energy_user}
\end{figure}
%%%%%%%%%%%%%%%%%%%%%%%%%%%%%%%%%%%%%%%%%%
%%%%%%%%%%%%%%%%%%%%%%%%%%%%%%%%%%%%%%%%%%
\begin{figure} [!t]
\centering \includegraphics [scale = 0.6] {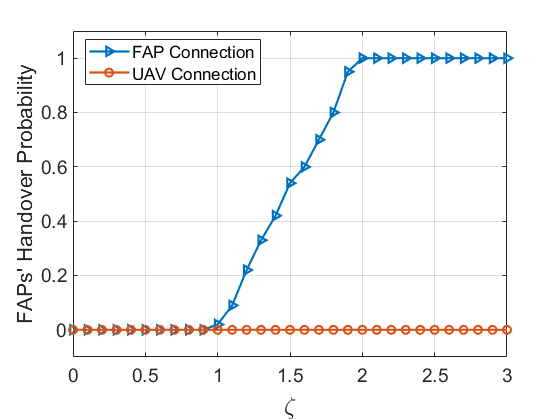}
\caption{\footnotesize The FAPs' handover probability versus different values of $\zeta=\dfrac{\upsilon}{\upsilon_{\rm{th}}}$.} \label{handover}
%\vspace{-.1in}
\end{figure}
%%%%%%%%%%%%%%%%%%%%%%%%%%%%%%%%%%%%%%%%%%
\vspace{.05in}
\noindent
\textit{Users' Access Delay through UAVs and UAVs' Energy Consumption:} We evaluate the users' access delay and the energy consumption of UAVs in Figs.~\ref{Delay_UAV} and~\ref{energy} when the ground user is located in both indoor and outdoor environments. As can be seen from Fig.~\ref{energy}, serving requests through UAVs, especially in such a case that ground users are located in the indoor environment, leads to consuming the energy of UAVs, calculated as follows~\cite{Sharma2019}
\begin{equation}
E_{u_k}^{(LoS)}(t)=L_c P_T(t) \tau_p +L_c P_R(t) \tau_p+P_j^{(LoS)}(t) (\tau_f-\tau_p),
\end{equation}
\begin{equation}
E_{u_k}^{(NLoS)}(t)=L_c P_T(t) \tau_p +L_c P_R(t) \tau_p+P_j^{(NLoS)}(t) (\tau_f-\tau_p),
\end{equation}
where $P_{T}(t)$ and $P_{R}(t)$ represent the power consumed for transmission and reception powers of $1$ Mb file, respectively. Moreover, $P_j(t)$, $\tau_f$, and $\tau_p$ denote the received power at ground user $GU_j$, and the flyby and the pause times of UAV $u_k$, respectively. On the other hand, it can be shown from Fig.~\ref{Delay_UAV} that the indoor users being served through UAVs, experience higher delay in comparison with outdoor users. Consequently, it can be seen that serving indoor users by UAVs could not be efficient from the aspect of user's access delay and energy consumption of UAVs. For this reason,  ground users in indoor areas are served by FAPs in inter-clusters.

Finally, Figs.~\ref{energy_user} and~\ref{handover} illustrate the advantage of serving outdoor users with both FAPs and UAVs in the CCUF framework. Fig.~\ref{energy_user} compares the average normalized UAVs' energy consumption in different scenarios, where $\psi$ illustrates the ratio of requests served by UAVs to the whole outdoor users' requests. For instance, $\psi=1$ means all requests in an outdoor environment are managed by UAVs regardless of the user's velocity, while in $\psi=0.7$, it is assumed that $70\%$ of outdoor users are HSUs, who are supported by UAVs and $30\%$ of users are LSUs, managed by FAPs. As it can be seen from Fig.~\ref{energy_user}, serving LSUs' requests by FAPs leads to a reduction in UAVs' energy consumption. Considering the fact that UAVs are limited  energy  caching  nodes, expanding  UAVs' lifetime is of paramount importance.
Moreover, Fig.~\ref{handover} illustrates the effect of users' velocity on the FAP's handover probability, where $\zeta=\dfrac{\upsilon}{\upsilon_{\rm{th}}}$. For comparison purposes, two scenarios are defined, where \textit{FAP Connection} is associated with a case that all outdoor users regardless of their velocities are supported by FAPs, while in \textit{UAV Connection}, LSUs and HSUs are supported by FAPs and UAVs, respectively. In this case, handover is triggered if the ground user leaves the current FAP's coverage before completely downloading one segment. As it can be seen from Fig.~\ref{handover}, serving HSUs ($\zeta>1$) leads to triggering frequent handovers, where the handover probability is one for $\zeta>2$, while there would not be any FAPs' handover by serving HSUs by UAVs.

%OOOOOOOOOOOOOOOOOOOOOOOOOOOOOOOOOOOOOOOOOOOOOOOOOOOOOOOOO
\vspace{-.1in}
\section{Conclusions} \label{sec:con}
%OOOOOOOOOOOOOOOOOOOOOOOOOOOOOOOOOOOOOOOOOOOOOOOOOOOOOOOOO
In this paper, we developed a Cluster-centric and Coded UAV-aided Femtocaching (CCUF) framework for an integrated and dynamic cellular network to maximize the number of requests served by caching nodes. To increase the cache diversity and to store distinct segments of contents in neighboring FAPs, we employed a two-phase clustering technique for FAPs' formation and UAVs' deployment. In this case, we formulated the success probability of the proposed CCUF framework. Moreover, in the cluster-centric cellular network, multimedia contents were coded based on their popularity profiles. In order to benefit the Coordinated Multi-Point (CoMP) technology and to improve the inter-cell interference, we determined the best value of the number of contents that should be stored completely. According  to  the  simulation results and by considering the best value of $\alpha$, the proposed CCUF framework results in an increase in the cache-hit-ratio, SINR,  and  cache  diversity  and decrease users’ access delay and cache redundancy. Going forward, several directions deserve further investigation. First, it is of interest to introduce a Reinforcement Learning (RL)-based method for outdoor environment, where ground users can be autonomously served by UAVs or FAPs, based on the dynamic population of their current locations and their speeds. Second, the optimum number of ground users to be served by a UAV in the proposed network needs to be analyzed.

%OOOOOOOOOOOOOOOOOOOOOOOOOOOOOOOOOOOOOOOOOOOOOOOOOOOOOOOOO
\bibliography{strings,refs}

\bibliographystyle{IEEEtran}
\bibliography{keylatex}

\begin{thebibliography}{00}

\bibitem{Chamola2020}
V. Chamola, V. Hassija, V. Gupta and M. Guizani,
\newblock ``A Comprehensive Review of the COVID-19 Pandemic and the Role of IoT, Drones, AI, Blockchain, and 5G in Managing its Impact,''
\newblock {\em IEEE Access}, vol. 8, pp. 90225-90265, May 2020.

%\bibitem{Hui2020}
%H. Hui, Y. Ding, Q. Shi, F. Li, Y. Song, and J. Yan,
%\newblock ``5G network-based Internet of Things for Demand Response in Smart Grid: A survey on Application Potential,''
%\newblock {\em  Applied Energy}, vol. 257, pp. 113972-113987, Jan. 2020.

\bibitem{Hajiakhondi2019}
Z.~Hajiakhondi-Meybodi, J.~Abouei, and A.~H.~F.~Raouf,
\newblock ``Cache Replacement Schemes Based on Adaptive Time Window for Video on Demand Services in Femtocell Networks,''
\newblock {\em IEEE Transactions on Mobile Computing}, vol. 18, no. 7, pp. 1476-1487, July 2019.

\bibitem{Hajiakhondi2020}
Z. HajiAkhondi-Meybodi, J. Abouei, M. Jaseemuddin and A. Mohammadi,
\newblock ``Mobility-Aware Femtocaching Algorithm in D2D Networks Based on Handover,''
\newblock {\em IEEE Transactions on Vehicular Technology}, vol. 69, no. 9, pp. 10188-10201, June 2020.

\bibitem{Jiang2019}
B.~Jiang, J.~Yang, H.~Xu, H.~Song, and G.~Zheng,
\newblock ``Multimedia Data Throughput Maximization in Internet-of-Things System Based on Optimization of Cache-Enabled UAV,''
\newblock {\em IEEE Internet of Things Journal}, vol. 6, no. 2, pp. 3525-3532, Apr. 2019.

\bibitem{Li2019}
B.~Li, Z.~Fei, and Y.~Zhang,
\newblock ``UAV Communications for 5G and Beyond: Recent Advances and Future Trends,''
\newblock {\em IEEE Internet of Things Journal}, vol. 6, no. 2, pp. 2241-2263, Apr. 2019.

\bibitem{Cheng2019}
F.~Cheng, G.~Gui, N.~Zhao, Y.~Chen, J.~Tang, and H.~Sari,
\newblock ``UAV-Relaying-Assisted Secure Transmission With Caching,''
\newblock {\em IEEE Transactions on Communications}, vol. 67, no. 5, pp. 3140-3153, May 2019.

\bibitem{Zhao2018}
N.~Zhao, F.~Cheng, F.~R. Yu, J.~Tang, Y.~Chen, G.~Gui, and H.~Sari,
\newblock ``Caching UAV Assisted Secure Transmission in Hyper-Dense Networks Based on Interference Alignment,''
\newblock {\em IEEE Transactions on Communications}, vol. 66, no. 5, pp. 2281-2294, May 2018.

\bibitem{Sharma2019}
V.~Sharma, I.~You, D.~N.~K. Jayakody, D.~G. Reina, and K.~K.~R. Choo,
\newblock ``Neural-Blockchain-Based Ultrareliable Caching for Edge-Enabled UAV Networks,''
\newblock {\em IEEE Transactions on Industrial Informatics}, vol. 15, no. 10, pp. 5723-5736, Oct. 2019.

%Deployment
\bibitem{Wang2019}
Z. Wang, L. Duan and R. Zhang,
\newblock ``Adaptive Deployment for UAV-Aided Communication Networks,''
\newblock {\em IEEE Transactions on Wireless Communications}, vol. 18, no. 9, pp. 4531-4543, Sept. 2019.

\bibitem{Liu2019}
X. Liu, Y. Liu and Y. Chen, \newblock ``Reinforcement Learning in Multiple-UAV Networks: Deployment and Movement Design,''
\newblock {\em IEEE Transactions on Vehicular Technology}, vol. 68, no. 8, pp. 8036-8049, Aug. 2019.


%Trajectrory
\bibitem{Samir2019}
M.~Samir, S.~Sharafeddine, C.~Assi, T.~M. Nguyen, and A.~Ghrayeb,
\newblock ``Trajectory Planning and Resource Allocation of Multiple UAVs for Data Delivery in Vehicular Networks,''
\newblock {\em IEEE Networking Letters}, vol. 1, no. 3, pp. 107-110, Sept. 2019.

\bibitem{Li2020_2}
S. Li, B. Duo, X. Yuan, Y. Liang and M. Di Renzo,
\newblock ``Reconfigurable Intelligent Surface Assisted UAV Communication: Joint Trajectory Design and Passive Beamforming,''
\newblock {\em IEEE Wireless Communications Letters}, vol. 9, no. 5, pp. 716-720, May 2020.


%Resource Allocation
\bibitem{Chen2019}
M.~Chen, W.~Saad, and C.~Yin,
\newblock ``Liquid State Machine Learning for Resource and Cache Management in LTE-U Unmanned Aerial Vehicle (UAV) Networks,''
\newblock {\em IEEE Transactions on Wireless Communications}, vol. 18, no. 3, pp. 1504-1517, Mar. 2019.

\bibitem{Yang2019}
Z. Yang, C. Pan, K. Wang and M. Shikh-Bahaei,
\newblock ``Energy Efficient Resource Allocation in UAV-Enabled Mobile Edge Computing Networks,''
\newblock {\em IEEE Transactions on Wireless Communications}, vol. 18, no. 9, pp. 4576-4589, Sept. 2019.

%Energy
\bibitem{Ji2019}
B.~Ji, Y.~Li, B.~Zhou, C.~Li, K.~Song, and H.~Wen,
\newblock ``Performance Analysis of UAV Relay Assisted IoT Communication Network Enhanced With Energy Harvesting,''
\newblock {\em IEEE Access}, vol. 7, pp. 38738-38747, Mar. 2019.


\bibitem{Zhang2018}
L.~Zhang, Z.~Zhao, Q.~Wu, H.~Zhao, H.~Xu, and X.~Wu,
\newblock ``Energy-Aware Dynamic Resource Allocation in UAV Assisted Mobile Edge Computing Over Social Internet of Vehicles,''
\newblock {\em IEEE Access}, vol. 6, pp. 56700-56715, Oct. 2018.

\bibitem{Suman2018}
S. Suman, S. Kumar and S. De,
\newblock ``Path Loss Model for UAV-Assisted RFET,''
\newblock {\em IEEE Communications Letters}, vol. 22, no. 10, pp. 2048-2051, Oct. 2018.

\bibitem{Hu2018_2}
Z. Hu, Z. Zheng, L. Song, T. Wang and X. Li,
\newblock ``UAV Offloading: Spectrum Trading Contract Design for UAV-Assisted Cellular Networks,''
\newblock {\em IEEE Transactions on Wireless Communications}, vol. 17, no. 9, pp. 6093-6107, Sept. 2018.


\bibitem{Lee2019}
J. Lee, K. Kim, M. Kim, J. Park, Y. K. Yoon and Y. J. Chong,
\newblock ``Measurement-Based Millimeter-Wave Angular and Delay Dispersion Characteristics of Outdoor-to-Indoor Propagation for 5G Millimeter-Wave Systems,''
\newblock {\em IEEE Access}, vol. 7, pp. 150492-150504, Oct. 2019.

\bibitem{Avanzato2020}
R. Avanzato and F. Beritelli,
\newblock ``A Smart UAV-Femtocell Data Sensing System for Post-Earthquake Localization of People,''
\newblock {\em  IEEE Access}, vol. 8, pp. 30262-30270, Feb. 2020.

\bibitem{Recayte2018}
E.~Recayte, F.~Lazaro, and G.~Liva,
\newblock ``Caching at the Edge with Fountain Codes,''
\newblock in Proc. {\em Advanced Satellite Multimedia Systems Conference and the Signal Processing for Space Communications Workshop (ASMS/SPSC)}, Berlin, Oct. 2018, pp. 1-6.

\bibitem{Ko2019}
D.~Ko, B.~Hong, and W.~Choi,
\newblock ``Probabilistic Caching Based on Maximum Distance Separable Code in a User-Centric Clustered Cache-Aided Wireless Network,''
\newblock {\em IEEE Transactions on Wireless Communications}, vol. 18, no. 3, pp. 1792-1804, Mar. 2019.

\bibitem{Shanmugam2013}
K. Shanmugam, N. Golrezaei, A. G. Dimakis, A. F. Molisch and G. Caire,
\newblock ``FemtoCaching: Wireless Content Delivery Through Distributed Caching Helpers,''
\newblock {\em IEEE Transactions on Information Theory}, vol. 59, no. 12, pp. 8402-8413, Dec. 2013.

\bibitem{Chen2017_2}
Z. Chen, J. Lee, T. Q. S. Quek and M. Kountouris,
\newblock ``Cooperative Caching and Transmission Design in Cluster-Centric Small Cell Networks,''
\newblock {\em IEEE Transactions on Wireless Communications}, vol. 16, no. 5, pp. 3401-3415, May 2017.

\bibitem{Liu2014}
A. Liu, and V. K. N. Lau, \newblock ``Cache-Enabled Opportunistic Cooperative MIMO for Video Streaming in Wireless Systems,''
\newblock {\em IEEE Transactions Signal Processing}, vol. 62, no. 2, pp. 390–402, Jan. 2014.

\bibitem{Yu2019}
Y. Yu, T. Hsieh, and A. Pang, \newblock ``Millimeter-Wave Backhaul Traffic Minimization for CoMP Over 5G Cellular Networks,''
\newblock {\em IEEE Transactions on Vehicular Technology}, vol. 68, no. 4, pp. 4003–4015, Apr. 2019.

\bibitem{Lin2020}
P. Lin, Q. Song, and A. Jamalipour, \newblock ``Multidimensional Cooperative Caching in CoMP-Integrated Ultra-dense Cellular Networks,''
\newblock {\em IEEE Transactions Wireless Communication}, vol. 19, no. 3, pp. 1977–1989, Mar. 2020.

\bibitem{Li2017}
H. Li, C. Yang, X. Huang, N. Ansari, and Z. Wang, \newblock ``Cooperative RAN Caching based on Local Altruistic Game for Single and Joint Transmissions,''
\newblock {\em IEEE Communication Letter}, vol. 21, no. 4, pp. 853–856, Apr. 2017.

\bibitem{Lin2019}
P. Lin, K. S. Khan, Q. Song, and A. Jamalipour, \newblock ``Caching in Heterogeneous Ultradense 5G Networks: A Comprehensive Cooperation Approach,''
\newblock {\em IEEE Vehicular Technology Magazine}, vol. 14, no. 2, pp. 22–32, Jun. 2019.

%\bibitem{Lin2020}
%P. Lin, Q. Song, J. Song, A. Jamalipour and F. R. Yu,
%\newblock ``Cooperative Caching and Transmission in CoMP-Integrated Cellular Networks Using Reinforcement Learning,''
%\newblock {\em IEEE Transactions on Vehicular Technology}, vol. 69, no. 5, pp. 5508-5520, May 2020.
\bibitem{Athukoralage2016}
D. Athukoralage, I. Guvenc, W. Saad and M. Bennis,
\newblock ``Regret Based Learning for UAV Assisted LTE-U/WiFi Public Safety Networks,''
\newblock {\em IEEE Global Communications Conference (GLOBECOM)}, Washington, Feb. 2017 pp. 1-7.

\bibitem{Zhu2020_2}
S. Zhu, L. Gui, N. Cheng, F. Sun and Q. Zhang,
\newblock ``Joint Design of Access Point Selection and Path Planning for UAV-Assisted Cellular Networks,''
\newblock {\em IEEE Internet of Things Journal}, vol. 7, no. 1, pp. 220-233, Jan. 2020.

\bibitem{Hajiakhondi2021}
Z. Hajiakhondi-Meybodi, A. Mohammadi and J. Abouei, \newblock ``Deep Reinforcement Learning for Trustworthy and Time-Varying Connection Scheduling in a Coupled UAV-Based Femtocaching Architecture,''
\newblock {\em IEEE Access}, vol. 9, pp. 32263-32281, Feb. 2021.

\bibitem{Afshang2016}
M. Afshang, H. S. Dhillon and P. H. J. Chong\newblock ``Fundamentals of Cluster-Centric Content Placement in Cache-Enabled Device-to-Device Networks,''
\newblock {\em IEEE Transactions on Communications}, vol. 64, no. 6, pp. 2511-2526, June 2016.

\bibitem{SZhang2018}
S. Zhang, P. He, K. Suto, P. Yang, L. Zhao and X. Shen, \newblock ``Cooperative Edge Caching in User-Centric Clustered Mobile Networks,''
\newblock {\em IEEE Transactions on Mobile Computing}, vol. 17, no. 8, pp. 1791-1805, Aug. 2018.

\bibitem{Xue2019}
K. Xue, L. Li, F. Yang, H. Zhang, X. Li and Z. Han,
\newblock ``Multi-UAV Delay Optimization in Edge Caching Networks: A Mean Field Game Approach,''
\newblock {\em Wireless and Optical Communications Conference}, Dec. 2019, pp. 1-5.

\bibitem{Zhang2021}
F. Zhang, G. Han, L. Liu, M. Martínez-García and Y. Peng, \newblock ``Joint Optimization of Cooperative Edge Caching and Radio Resource Allocation in 5G-Enabled Massive IoT Networks,''
 Accepted in \newblock {\em IEEE Internet of Things Journal}, Mar. 2021. 
 
\bibitem{Choi2006}
Y. Choi, C. S. Kim and S. Bahk,
\newblock ``Flexible Design of Frequency Reuse Factor in OFDMA Cellular Networks,''
\newblock {\em IEEE International Conference on Communications}, Istanbul, 2006, pp. 1784-1788. 
 
\bibitem{Wireless}	
Wireless Communication. Available online: http://www.wirelesscommunication.nl/ reference/chaptr04/cellplan/reuse.html (accessed on 21-06-11).
 
\bibitem{Al-Samman2019}
A.~M. Al-Samman, T.~A. Rahman, T.~Al-Hadhrami, A.~Daho, M.~N. Hindia, M.~H.
  Azmi, K.~Dimyati, and M.~Alazab,
\newblock ``Comparative Study of Indoor Propagation Model Below and Above 6 GHz for 5G Wireless Networks,''
\newblock {\em Electronics}, vol. 18, no. 1, Jan. 2019.

\bibitem{Kanungo2002}
T. Kanungo, D. M. Mount, N. S. Netanyahu, C. D. Piatko, R. Silverman and A. Y. Wu,
\newblock ``An efficient k-means clustering algorithm: analysis and implementation,''
\newblock {\em IEEE Transactions on Pattern Analysis and Machine Intelligence}, vol. 24, no. 7, pp. 881-892, July 2002. 
 
\bibitem{Albertsen2019}
C. M. Albertsen,
\newblock ``Generalizing the First-Difference Correlated Random Walk for Marine Animal Movement Data,''
\newblock {\em Scientific Reports}, vol. 9, no. 1, pp. 4017--4031, Mar. 2019.

\bibitem{Hajiakhondi2020_1}
Z.~HajiAkhondi-Meybodi, M.~S.~Beni, K. N. Plataniotis, and A. Mohammadi
\newblock ``Bluetooth Low Energy-based Angle of Arrival Estimation via Switch Antenna Array for Indoor Localization,''
\newblock {\em  International Conference on Information Fusion}, July 2020.

\bibitem{Hajiakhondi2020_2}
Z.~HajiAkhondi-Meybodi, M.~S.~Beni, A. Mohammadi, and K. N. Plataniotis,
\newblock ``Bluetooth Low Energy-based Angle of Arrival Estimation in Presence of Rayleigh Fading,''
\newblock {\em  IEEE International Conference on Systems, Man, and Sybernetics}, 2020.











%\bibitem{Chen2019_2}
%M.~Chen, W.~Saad, and C.~Yin,
%\newblock ``Echo-Liquid State Deep Learning for 360 Content Transmission and Caching in Wireless VR Networks With Cellular-Connected UAVs,''
%\newblock {\em IEEE Transactions on Communications}, vol. 67, no. 9, pp. 6386-6400, Sept. 2019.

%\bibitem{Wu2018}
%H.~Wu, X.~Tao, N.~Zhang, and X.~Shen,
%\newblock ``Cooperative UAV Cluster-Assisted Terrestrial Cellular Networks for Ubiquitous Coverage,''
%\newblock {\em IEEE Journal on Selected Areas in Communications}, vol. 36, no. 9, pp. 2045-2058, Sept. 2018.

%\bibitem{Chen2017}
%M.~Chen, M.~Mozaffari, W.~Saad, C.~Yin, M.~Debbah, and %C.~S. Hong,
%\newblock ``Caching in the Sky: Proactive Deployment of Cache-Enabled Unmanned Aerial Vehicles for Optimized Quality-of-Experience,''
%\newblock {\em IEEE Journal on Selected Areas in Communications}, vol. 35, no. 5, pp. 1046-1061, May 2017.





%\bibitem{Ren2019}
%D. Ren, X. Gui, K. Zhang, and J. Wu, J.,
%\newblock ``Hybrid Collaborative Caching in Mobile Edge Networks: An Analytical Approach,''
%\newblock {\em Computer Networks}, vol. 158, pp.1-16, July 2019.





%\bibitem{Lin2020}
%P. Lin, Q. Song, J. Song, A. Jamalipour, and F. R. Yu,
%\newblock ``Cooperative Caching and Transmission in CoMP-Integrated Cellular Networks Using Reinforcement Learning,''
%\newblock {\em IEEE Transactions on Vehicular Technology}, vol. 69, no. 5, pp. 5508-5520, Mar. 2020.


%\bibitem{Qiu2019_2}
%L.~Qiu and G.~Cao,
%\newblock ``Popularity-Aware Caching Increases the Capacity of Wireless Networks,''
%\newblock {\em IEEE Transactions on Mobile Computing}, vol.19, no. 1, pp. 173--87, Jan. 2019.







%\bibitem{Zhang2020}
%T. Zhang, Y. Wang, Y. Liu, W. Xu and A. Nallanathan,
%\newblock ``Cache-enabling UAV Communications: Network %Deployment and Resource Allocation,''
%\newblock {\em IEEE Transactions on Wireless Communications}, July 2020.

\end{thebibliography}
\begin{IEEEbiography}[{\includegraphics[width=1in,height=1.25in,clip,keepaspectratio]{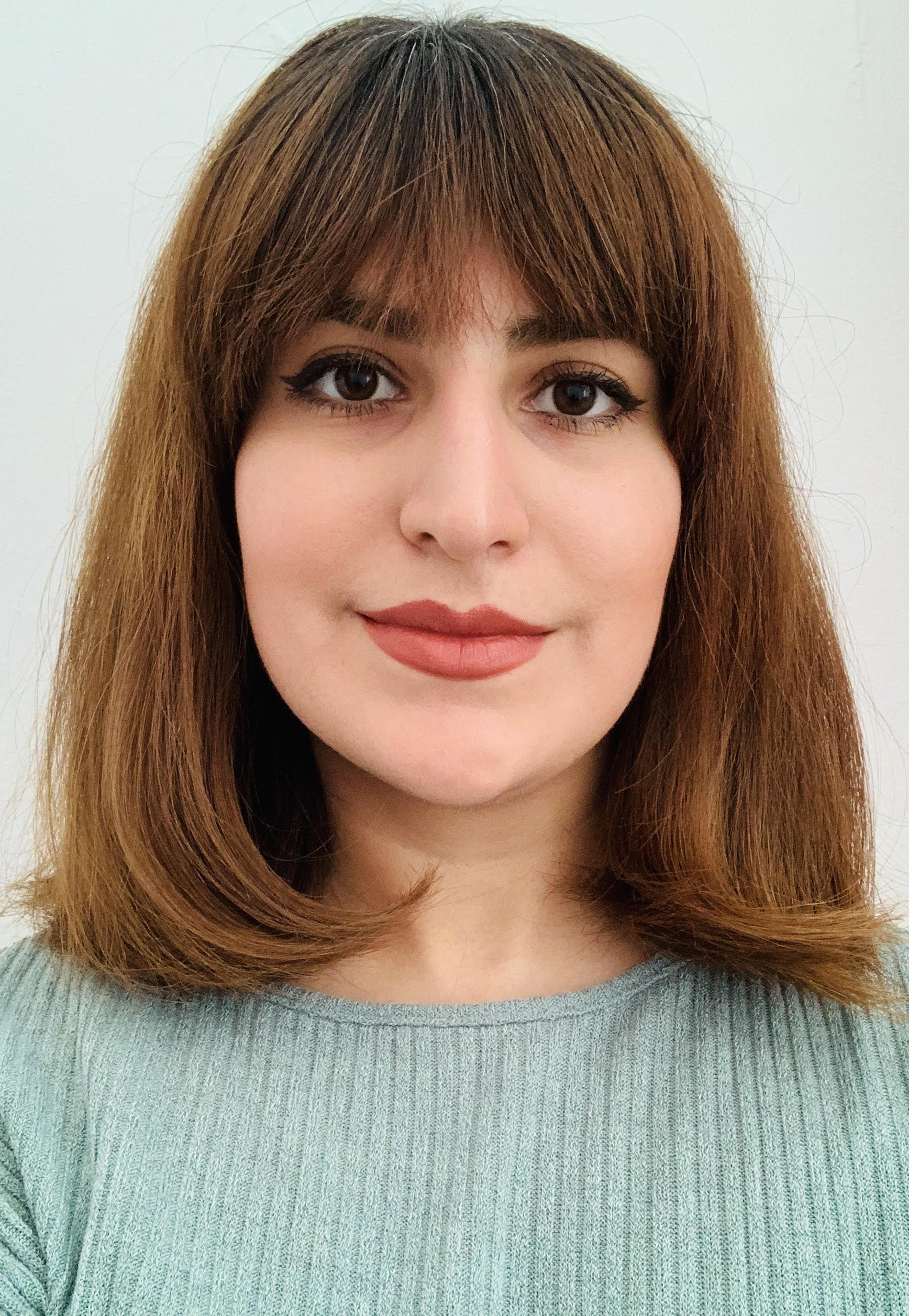}}]
{Zohreh Hajiakhondi-Meybodi} received the B.Sc. degree in Communication Engineering from Yazd University, Yazd, Iran and the M.Sc. degree in Communication Systems Engineering (with the highest honor) from Yazd University, Yazd, Iran in 2013 and 2017, respectively. She is a Ph.D. degree candidate at Electrical and Computer Engineering (ECE), Concordia University, Montreal, Canada. Since 2019, she has been an active member of I-SIP Lab at Concordia University. Her research interests include general areas of wireless communication networks with a particular emphasis on Femtocaching, Internet of Things (IoT), Indoor Localization, Optimization Algorithms, and Multimedia Wireless Sensor Networks (WMSN).
\end{IEEEbiography}
\begin{IEEEbiography}[{\includegraphics[width=1in,height=1.25in,clip,keepaspectratio]{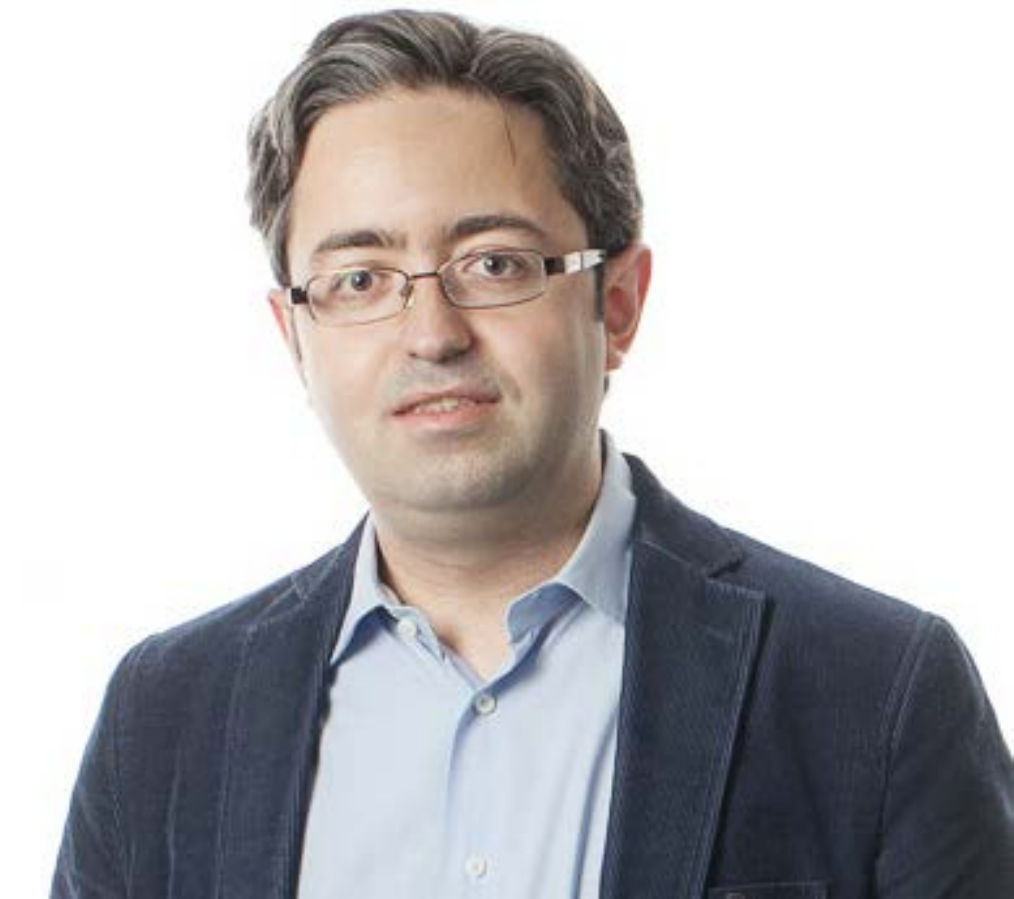}}]
{Arash Mohammadi}  (S'08-M'14-SM'17) is currently an Associate Professor with Concordia Institute for Information Systems Engineering, Concordia University, Montreal, QC, Canada. Prior to joining Concordia University and for 2 years, he was a Postdoctoral Fellow with the Department of Electrical and Computer Engineering, University of Toronto, Toronto, ON, Canada. Dr. Mohammadi is a registered professional engineer in Ontario. He is Director-Membership Developments of  IEEE Signal Processing Society (SPS); General Co-Chair of ``2021 IEEE International Conference on Autonomous Systems (ICAS),'' and; Guest Editor for IEEE Signal Processing Magazine (SPM) Special Issue on ``Signal Processing for Neurorehabilitation and Assistive Technologies''. He is also currently serving as Associate Editor on the editorial board of IEEE Signal Processing Letters. He was Co-Chair of ``Symposium on Advanced Bio-Signal Processing and Machine Learning for Assistive and Neuro-Rehabilitation Systems'' as part of 2019 IEEE GlobalSIP, and ``Symposium on Advanced Bio-Signal Processing and Machine Learning for Medical Cyber-Physical Systems,'' as a part of IEEE GlobalSIP'18; The Organizing Chair of 2018 IEEE Signal Processing Society Video and Image Processing (VIP) Cup, and the Lead Guest  Editor for IEEE Transactions on Signal \& Information Processing over Networks Special Issue on ``Distributed Signal Processing for Security and Privacy in Networked Cyber-Physical Systems''. He is recipient of several distinguishing awards including the Eshrat Arjomandi Award for outstanding Ph.D. dissertation from Electrical Engineering and Computer Science Department, York University, in 2013; Concordia President's Excellence in Teaching Award in 2018, and; 2019 Gina Cody School of Engineering and Computer Science’s Research and Teaching awards in the new scholar category.
\end{IEEEbiography}

\begin{IEEEbiography}[{\includegraphics[width=1in,height=1.25in,clip,keepaspectratio]{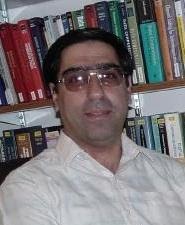}}]
{Jamshid Abouei} (S05, M11, SM13) received the B.Sc. degree in Electronics Engineering and the M.Sc. degree in Communication Systems Engineering (with the highest honor) both from Isfahan University of Technology (IUT), Iran, in 1993 and 1996, respectively, and the Ph.D. degree in Electrical Engineering from University of Waterloo, Canada, in 2009. He joined with the Department of Electrical Engineering, Yazd University, Iran, in 1996 (as a Lecturer) and was promoted to Assistant Professor in 2010, and Associate Professor in 2015. From 1998 to 2004, he served as a Technical Advisor and Design Engineer in the R \& D Center and Cable Design Department in SGCC, Iran. From 2009 to 2010, he was a Postdoctoral Fellow in the Multimedia Lab, in the Department of Electrical \& Computer Engineering, University of Toronto, Canada, and worked as a Research Fellow at the Self-Powered Sensor Networks (ORF-SPSN) consortium. During his sabbatical, he was an Associate Researcher in the Department of Electrical, Computer and Biomedical Engineering, Ryerson University, Toronto, Canada. Dr Abouei was the International Relations Chair in 27th ICEE2019 Conference, Iran, in 2019. Currently, Dr Abouei directs the research group at the Wireless Networking Laboratory (WINEL), Yazd University, Iran. His research interests are in the next generation of wireless networks (5G) and wireless sensor networks (WSNs), with a particular emphasis on PHY/MAC layer designs including the energy efficiency and optimal resource allocation in cognitive cell-free massive MIMO networks, multi-user information theory, mobile edge computing and femtocaching. Dr Abouei is a Senior IEEE member and a member of the IEEE Information Theory. He has received several awards and scholarships, including FOE and IGSA awards for excellence in research in University of Waterloo, Canada, MSRT Ph.D. Scholarship from the Ministry of Science, Research and Technology, Iran in 2004, Distinguished Researcher award in province of Yazd, Iran, 2011, and Distinguished Researcher award in Electrical Engineering Department, Yazd University, Iran, 2013. He is a recipient of the best paper award for the IEEE Iranian Conference on Electrical Engineering (ICEE 2018).
\end{IEEEbiography}

\begin{IEEEbiography}[{\includegraphics[width=1in,height=1.25in,clip,keepaspectratio]{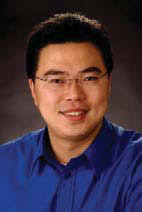}}]{Ming Hou} (M'05-SM'07) is currently a Senior Defence Scientist with Defence Research and Development Canada (DRDC) and the Principal Authority of Human-Technology Interactions with the Department of National Defence (DND), Canada. He is responsible for providing science-based advice at national and international levels to the Canadian Armed Forces (CAF) and coalition partners about the investment in and application of advanced technologies for human–machine systems requirements. He is an Integrator for the Canadian government 16 billion IDEaS program and one of the three Scientific Advisors to the Canadian National Centre of Expertise in Human Systems Performance with responsibilities for guiding national research and development activities in automation, robotics, and telepresence. He also gives advice for the development of National Defence AI Science and Technology Strategy and Roadmap to the CAF and DND. He is the Co-Chair of Human Factors Specialist Team within NATO Joint Capability Group on Unmanned Aircraft Systems (UAS). His book Intelligent Adaptive Systems: An Interaction-Centered Design Perspective became a guiding document to the development of NATO Standard Recommendations on UAS Human Systems Integration Guide book, UAS Human Factors Experimentation Guidelines and UAS Sense and Avoid Guidance. As one of the four invited lecturers, he delivers NATO Lecture Series on UAVs: Technological Challenges, Concepts of Operations, and Regulatory Issues. He also serves for multiple international associations/programs as a chair and a board members.\\
\end{IEEEbiography}

\begin{IEEEbiography}[{\includegraphics[width=1in,height=1.25in,clip,keepaspectratio]{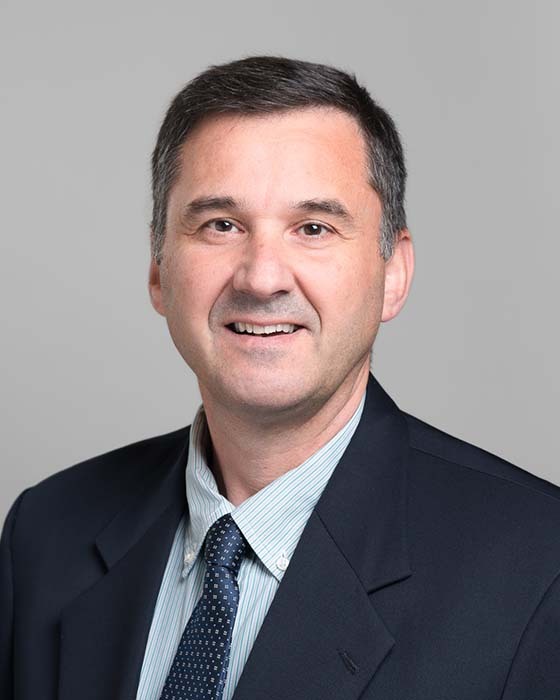}}]
{Konstantinos N. (Kostas) Plataniotis} is a Professor and the Bell Canada Chair in Multimedia with the ECE Department at the University of Toronto. He is the founder and inaugural Director-Research for the Identity, Privacy and Security Institute (IPSI) at the University of Toronto and he has served as the Director for the Knowledge Media Design Institute (KMDI) at the University of Toronto from January 2010 to July 2012. His research interests are: knowledge and digital media design, multimedia systems, biometrics, image \& signal processing, communications systems and pattern recognition. Among his publications in these fields are the books entitled “WLAN positioning systems’ (2012) and `Multi-linear subspace learning: Reduction of multidimensional data’ (2013). Dr. Plataniotis is a registered professional engineer in Ontario, Fellow of the IEEE and Fellow of the Engineering Institute of Canada. He has served as the Editor-in-Chief of the IEEE Signal Processing Letters, and as Technical Co-Chair of the IEEE 2013 International Conference in Acoustics, Speech and Signal Processing. He was the IEEE Signal Processing Society Vice President for Membership (2014 -2016). He is the General Co-Chair for 2017 IEEE GlobalSIP, the 2018 IEEE International Conference on Image Processing (ICIP 2018), and the 2021 IEEE International Conference on Acoustics, Speech and Signal Processing (ICASSP 2021).
\end{IEEEbiography}

\end{document}